\documentclass[useAMS,usenatbib]{mnras}

\usepackage{natbib}
\usepackage{tabu,booktabs}
\usepackage{amsmath}
\usepackage{xcolor}
\usepackage[english]{babel}
\usepackage{graphicx}
\usepackage{xspace}

\usepackage{longtable}
\usepackage{threeparttablex}
\usepackage{array}

\def \etal {et~al.~}

\newcommand{\hMpc}{{\ifmmode{h^{-1}{\rm Mpc}}\else{$h^{-1}$Mpc}\fi}}
\newcommand{\Mpc}{{\ifmmode{{\rm Mpc}}\else{Mpc}\fi}}
\newcommand{\hkpc}{{\ifmmode{h^{-1}{\rm kpc}}\else{$h^{-1}$kpc}\fi}}
\newcommand{\kpc}{{\ifmmode{ {\rm kpc} }\else{{\rm kpc}}\fi}}
\newcommand{\kms}{{\ifmmode{ {\rm km\,s^{-1}} }\else{ ${\rm km\,s^{-1}}$ }\fi}}
\newcommand{\hMsun}{{\ifmmode{h^{-1}{\rm {M_{\odot}}}}\else{$h^{-1}{\rm{M_{\odot}}}$}\fi}}
\newcommand{\Msun}{{\ifmmode{{\rm M}_{\odot}}\else{${\rm M}_{\odot}$}\fi}}
\newcommand{\Mhalo}{{\ifmmode{M_{\rm halo}}\else{$M_{\rm halo}$}\fi}}
\newcommand{\Rvir}{{\ifmmode{R_{\rm vir}}\else{$R_{\rm vir}$}\fi}}
\newcommand{\Mvir}{{\ifmmode{M_{\rm vir}}\else{$M_{\rm vir}$}\fi}}
\newcommand{\Mstar}{{\ifmmode{M_{\rm star}}\else{$M_{\rm star}$}\fi}}
\newcommand{\Vrot}{{\ifmmode{V_{\rm rot}}\else{$V_{\rm rot}$}\fi}}
\newcommand{\ltsima}{$\; \buildrel < \over \sim \;$}
\newcommand{\gtsima}{$\; \buildrel > \over \sim \;$}
\newcommand{\lsim}{\lower.5ex\hbox{\ltsima}}
\newcommand{\gsim}{\lower.5ex\hbox{\gtsima}}

\def\lesssim{\mathrel{\hbox{\rlap{\hbox{\lower4pt\hbox{$\sim$}}}\hbox{$<$}}}}
\def\gtrsim{\mathrel{\hbox{\rlap{\hbox{\lower4pt\hbox{$\sim$}}}\hbox{$>$}}}}

\newcommand{\beq}{\begin{equation}}
\newcommand{\eeq}{\end{equation}}
\def\beqa{\begin{eqnarray}}
\def\eeqa{\end{eqnarray}}
\def\LCDM{\ensuremath{\Lambda}CDM}

\def\head{ \vbox to 0pt{\vss \hbox to 0pt{\hskip 440pt\rm
      LA-UR-10-07069\hss} \vskip 25pt}}

\def \kms {\ifmmode  \,\rm km\,s^{-1} \else $\,\rm km\,s^{-1}  $ \fi }
\def \kpc {\ifmmode  {\rm kpc}  \else ${\rm  kpc}$ \fi  }  
\def \hkpc {\ifmmode  {h^{-1}\rm kpc}  \else ${h^{-1}\rm kpc}$ \fi  }  
\def \hMpc {\ifmmode  {h^{-1}\rm Mpc}  \else ${h^{-1}\rm Mpc}$ \fi  }  
\def \Mpch {\ifmmode  {h^{-1}\rm Mpc}  \else ${h^{-1}\rm Mpc}$ \fi  }  
\def \Msun {\ifmmode {\rm M}_{\odot} \else ${\rm M}_{\odot}$ \fi} 
\def \hMsun {\ifmmode h^{-1}\,\rm M_{\odot} \else $h^{-1}\,\rm M_{\odot}$ \fi}

\def \LCDM {\ifmmode \Lambda{\rm CDM} \else $\Lambda{\rm CDM}$ \fi}
\def \sig8 {\ifmmode \sigma_8 \else $\sigma_8$ \fi} 
\def \OmegaM {\ifmmode \Omega_{\rm m} \else $\Omega_{\rm m}$ \fi} 
\def \Omegab {\ifmmode \Omega_{\rm b} \else $\Omega_{\rm b}$ \fi} 
\def \OmegaL {\ifmmode \Omega_{\rm \Lambda} \else $\Omega_{\rm \Lambda}$\fi} 
\def \Deltavir {\ifmmode \Delta_{\rm vir} \else $\Delta_{\rm vir}$ \fi}
\def \rhocrit {\ifmmode \rho_{\rm crit} \else $\rho_{\rm crit}$ \fi}
\def \rhou {\ifmmode \rho_{\rm u} \else $\rho_{\rm u}$ \fi}
\def \zc {\ifmmode z_{\rm c} \else $z_{\rm c}$ \fi}

\def\aj{AJ}%
\def\araa{ARA\&A}%
\def\apj{ApJ}%
\def\apjl{ApJ}%
\def\apjs{ApJS}%
\def\ao{Appl.~Opt.}%
\def\aap{A\&A}%
\def\mnras{MNRAS}%
\def\pasp{PASP}%
\def\nat{Nature}%
\def\head{ .ps \vbox to 0pt{\vss \hbox to 0pt{\hskip 440pt\rm
      LA-UR-10-07069\hss} \vskip 25pt}} 

\def \spose#1{\hbox  to 0pt{#1\hss}}  
\def \lta{\mathrel{\spose{\lower 3pt\hbox{$\sim$}}\raise 2.0pt\hbox{$<$}}}
\def \gta{\mathrel{\spose{\lower 3pt\hbox{$\sim$}}\raise 2.0pt\hbox{$>$}}}

\title[Stellar disk structure]{NIHAO-UHD: The properties of MW-like stellar disks in high resolution cosmological simulations}

\author[T. Buck \etal] {Tobias Buck$^{1}$\thanks{E-mail: tbuck@aip.de},
Aura Obreja$^{2}$, Andrea V. Macci\`o$^{3,4}$, Ivan Minchev$^1$,
\newauthor{Aaron A. Dutton$^{3}$ \& Jeremiah P. Ostriker$^{5,6}$}\\
$^1$Leibniz-Institut f\"ur Astrophysik Potsdam (AIP), An der Sternwarte 16, D-14482 Potsdam, Germany\\
$^2$Universit\"ats-Sternwarte M\"unchen, Scheinerstraße 1, D-81679 M\"unchen, Germany\\
$^3$New York University Abu Dhabi, PO Box 129188, Saadiyat Island, Abu Dhabi, United Arab Emirates\\
$^4$Max-Planck-Institut f\"ur Astronomie, K\"onigstuhl 17, 69117 Heidelberg, Germany\\
$^5$Department of Astronomy, Columbia University, New York, NY 10027, USA\\
$^6$Department of Astrophysical Sciences, Princeton University, Princeton, NJ 08544, USA}

\begin{document}

\date{Accepted 2019 November 15. Received 2019 November 12; in original form 2019 September 9}

\pagerange{\pageref{firstpage}--\pageref{lastpage}} \pubyear{2019}

\maketitle

\label{firstpage}

\begin{abstract}
Simulating thin and extended galactic disks has long been a challenge in computational astrophysics. We introduce the NIHAO-UHD suite of cosmological hydrodynamical simulations of Milky Way mass galaxies and study stellar disk properties such as stellar mass, size and rotation velocity which agree well with observations of the Milky Way and local galaxies. In particular, the simulations reproduce the age-velocity dispersion relation and a multi-component stellar disk as observed for the Milky Way.
Half of our galaxies show a double exponential vertical profile, while the others are well described by a single exponential model which we link to the disk merger history. In all cases, mono-age populations follow a single exponential whose scale height varies monotonically with stellar age and radius. The scale length decreases with stellar age while the scale height increases. The general structure of the stellar disks is already set at time of birth as a result of the inside-out and upside-down formation. Subsequent evolution modifies this structure by increasing both the scale length and height of all mono-age populations. Thus, our results put tight constraints on how much dynamical memory stellar disks can retain over cosmological timescales. Our simulations demonstrate that it is possible to form thin galactic disks in cosmological simulations provided there are no significant stellar mergers at low redshifts. Most of the stellar mass is formed in-situ with only a few percent ($\lesssim5\%$) brought in by merging satellites at early times. Redshift zero snapshots and halo catalogues are publicly available.
\end{abstract}

\noindent
\begin{keywords}

Galaxy: structure --- galaxies: kinematics and dynamics --- Galaxy: disk --- galaxies:
  formation --- Galaxy: evolution --- methods: numerical

 \end{keywords}


\begin{table}
\begin{center}
\caption{Simulation parameters for the dark matter, gas and stellar particles in our simulation runs.}
\label{tab:res}
\begin{tabular}{c c c c}
		\hline\hline
		property & particle mass & Force soft. & particles within $R_{200}$\\
		  & [$10^5 \Msun$] & [pc] & \\
		\hline
        \multicolumn{4}{c}{g2.79e12}\\
        \hline
		DM & 5.141 & 620 & 5,413,017\\
		GAS & 0.938 & 265 & 2,224,231\\
		STARS & 0.313 & 265 & 8,249,934\\
        \hline
        \multicolumn{4}{c}{g1.12e12}\\
        \hline
        		DM & 1.523 & 414 & 7,454,484\\
		GAS & 0.278 & 177 & 3,041,329\\
		STARS & 0.093 & 177 & 11,087,938\\
		\hline
        \multicolumn{4}{c}{g8.26e11}\\
        \hline
        		DM & 2.168 & 466 & 3,742,713\\
		GAS & 0.396 & 199 & 1,665,173\\
		STARS & 0.132 & 199 & 4,171,413\\
        \hline
        \multicolumn{4}{c}{g7.55e11}\\
        \hline
        		DM & 1.523 & 414 & 4,980,547\\
		GAS & 0.278 & 177 & 2,618,668\\
		STARS & 0.093 & 177 & 4,773,743\\
	\hline
        \multicolumn{4}{c}{g7.08e11}\\
        \hline
        		DM & 1.11 & 373 & 4,439,242\\
		GAS & 0.203 & 159 & 2,004,656\\
		STARS & 0.068 & 159 & 4,861,964\\
        \hline
        \multicolumn{4}{c}{g6.96e11}\\
        \hline
        		DM & 1.523 & 414 & 4,020,317 \\
		GAS & 0.278 & 177 & 1,879,331 \\
		STARS & 0.093 & 177 & 2,714,213 \\
	\hline
\end{tabular}
\end{center}
\end{table}

\section{Introduction} \label{sec:introduction}

Studying the formation and evolution of disk galaxies like the Milky Way (MW) or M31 has a long history in theoretical astrophysics and by now a basic formation scenario has been established \citep{White1978,Fall1980,Mo1998,Dutton2007}. Despite the great effort spent on simulating the formation of disk galaxies from cosmological initial conditions, the details of their formation mechanisms are still unclear. Early works on computational galaxy formation suffered from the ``over-cooling" problem and angular momentum losses resulted in compact bulge dominated systems \citep{Katz1991,Navarro2000}.

Since then it has been realised that the inclusion of stellar feedback, e.g. from exploding supernovae, is the key ingredient in shaping disk galaxies by effectively removing low angular momentum gas from the central galactic regions \citep{Governato2010,Brook2011}. 

By now many studies report successes in reproducing key observables of disk galaxies \citep[e.g.][]{Aumer2013,Stinson2013b,Marinacci2014,Wang2015,Grand2017,Hopkins2018} due to the implementation of efficient stellar feedback which can be injected as thermal \citep[e.g.][]{Stinson2006,Keller2014} and/or kinetic energy  \citep[e.g.][]{Springel2003,Oppenheimer2006,DallaVecchia2008} in surrounding gas cells or particles. 
However, it has been found that supernova feedback alone is not capable of regulating star formation in MW-type galaxies. \citet{Stinson2013b} and \citet{Aumer2013} advocated that early stellar feedback from young massive stars is critical in preventing the early onset of star formation and thus minimizing the central growth of the bulge component. Other forms of feedback studied in the literature include radiation pressure \citep[e.g.][]{Hopkins2012,Sparre2015,Rosdahl2015}, cosmic rays \citep[e.g.][]{Pakmor2016,Pfrommer2017} or Active Galactic Nuclei (AGN) feedback \citep[e.g.][]{Springel2005,Sijacki2007,Dubois2013,Weinberger2017}. The relative importance of the hydro solver is minimal in comparison to different feedback implementations \citep[see e.g. the Aquila comparison project][]{Scannapieco2012}. 

Further complications arise from the choice of parameters for a given feedback scheme. Even at the highest resolutions achievable by state-of-the-art simulations the stellar particles used to trace the stellar density field still represent simple stellar populations of thousands of stars characterised by an initial mass function (IMF), while local processes in the ISM are barely resolved. Thus all current galaxy formation models rely on parametrised models where the choice of parameters values varies greatly. Recently, the impact of systematically varying galaxy formation parameters has been studied in detail \citep[e.g.][]{Dutton2019,Munshi2019} and their strong influences on galaxy properties such as the structure of the dark matter halo could be established. For example, systematically varying the density threshold for star formation results in vastly different morphologies of the young stars; \citet{Buck2019} used this fact to observationally put constraints on this particular parameter. In general, all current galaxy formation models suffer from uncertainties due an (arbitrary) choice of parameters which in the worst case might need to be re-tuned when changing the resolution of the simulation. 

\begin{figure}
\begin{center}
\includegraphics[width=\columnwidth]{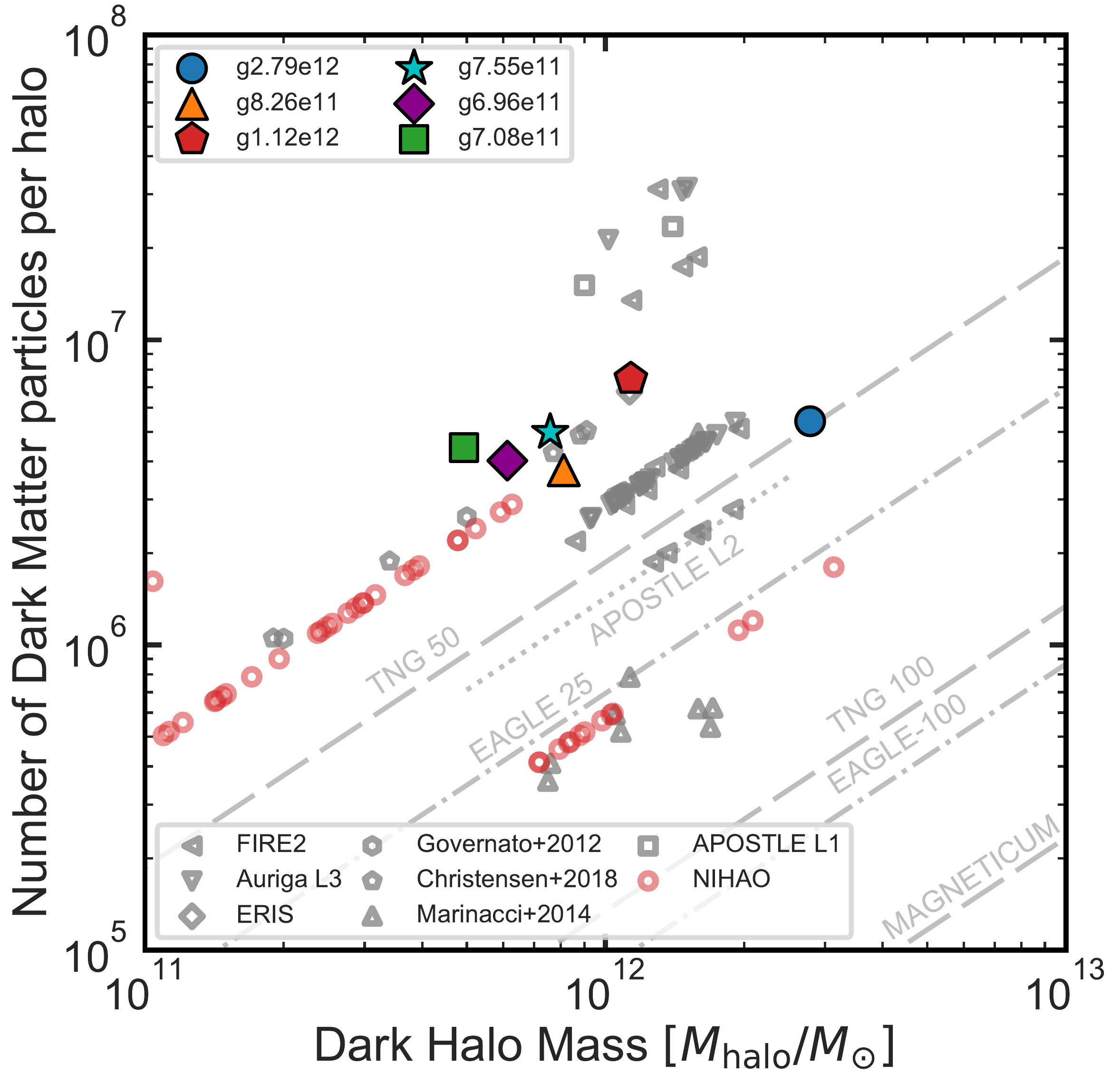}
\end{center}
\vspace{-.35cm}
\caption{Number of dark matter particles per halo vs. halo mass for the six NIHAO-UHD simulations (big colored symbols with black contour) in comparison to state-of-the-art cosmological simulations in the literature (gray symbols) and a selection of the original NIHAO galaxies (red circles).}
\label{fig:mass_res}
\end{figure}

\begin{figure*}
\begin{center}
\includegraphics[width=.9\textwidth]{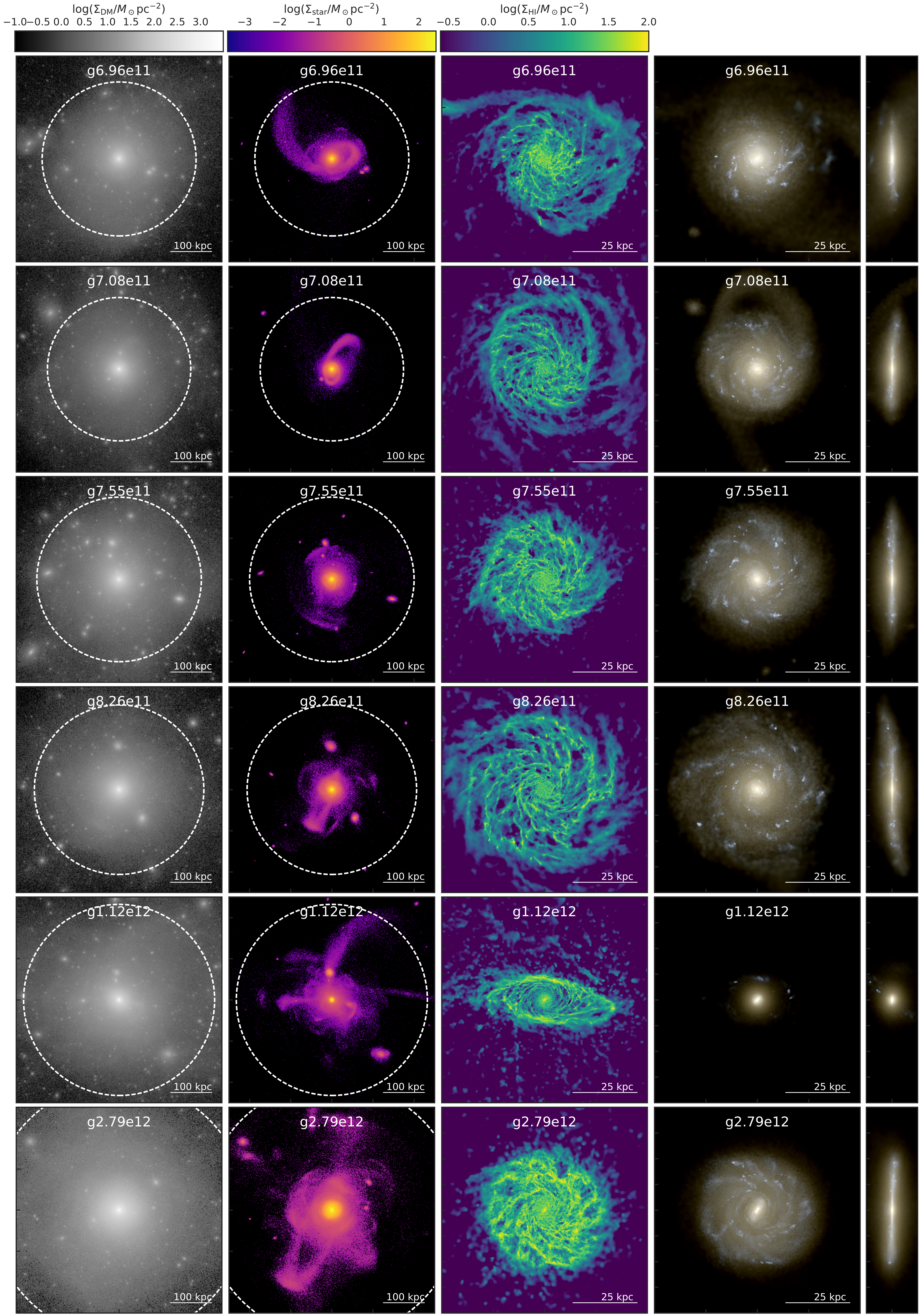}
\end{center}
\vspace{-.35cm}
\caption{Surface density maps of the galaxies in face-on views. From left to right: dark matter, stars, HI gas and face- and edge-on RGB images of the stars in the wavelength bands $g,r,i$. In the left most columns the dashed white circle indicates the virial radius $R_{200}$. For this figure, we calculate the HI fraction following the fitting formulas of \citet{Rahmati2013}.}
\label{fig:impression}
\end{figure*}

Here, we introduce the NIHAO-UHD\footnote{where UHD stands for Ultra High Definition} simulations, higher resolution versions of galaxies taken from the Numerical Investigation of a Hundred Astronomical Objects (NIHAO) simulation suite \citep{Wang2015}. The NIHAO-UHD galaxies are a set of six high-resolution zoom-in simulations of the formation of MW-like galaxies sharing exactly the same initial conditions as their lower resolution (roughly a factor of 8 in mass) NIHAO counterparts and are run with exactly the same feedback parameters. In this paper, we show that the physically motivated NIHAO feedback model is robust across resolution levels. Together with the AURIGA galaxies \citep{Grand2017} and the FIRE2 sample of MW-like galaxies \citep{Garrison-Kimmel2018}, the NIHAO-UHD sample thus comprises one of the largest and highest resolution sample \citep[still a factor of $\sim5$ higher mass resolution than Illustris-TNG50,][]{Pillepich2019} of zoom-in MW-like simulations in the literature (see figure \ref{fig:mass_res}). 

While parts of the NIHAO-UHD sample have been used to study the satellite population of MW like galaxies \citep{Buck2019} or its central stellar bar \citep{Buck2018,Buck2019b}, in this paper we focus on the general properties of the central disk galaxies and establish the excellent agreement with observational results for the MW in particular and local disk galaxies from the SPARC sample \citep{Lelli2016} in general. In particular, we use this sample of high-resolution disk galaxies to study the structure and build-up of their present-day stellar disks.

This paper is organised as follows: We start by introducing the set of simulations and the numerical method in \S2. In \S3 we show general galaxy properties such as stellar mass growth, stellar disk sizes, rotation curves or the age velocity dispersion relation in comparison to observational data to establish the realism of our galaxies. From there we analyse and discuss the detailed structure of the stellar disk in  \S4. We focus hereby on the  disk's vertical structure as a function of stellar age and leave the study of the chemical abundance patterns for a future paper \citep{Buck2019c}. We conclude this work in \S5 with a summary and discussion of our results.

\begin{table*}
\label{tab:props}
\begin{center}
\caption{Simulation properties of the main galaxies: \normalfont{We state the total virial mass $M_{200}$, the virial radius $R_{200}$ and the total stellar mass, $M_{\rm star}$, as well as the total amount of gas, $M_{\rm gas}$, within the virial radius. We further report the disk scale length, $R_{\rm d}$, the bulge effective radius, $R_{\rm eff}$, the bulge S\'ersic index, $n$, the disk-to-total ratio D/T, as well as the scale height of the thin and thick stellar disk at the solar radius ($8\pm1$ kpc). The last column contains the circular velocity measured at $2.2R_{\rm d}$.}}
\begin{tabular}{l c c c c c c c c c c c}
		\hline\hline
		simulation & $M_{200}$ & $R_{200}$ & $M_{\rm star}$ & $M_{\rm gas}$ & $R_{\rm d}$ & $R_{\rm eff}$ & $n$ & D/T & $h_{\rm z}^{\rm t}$ & $h_{\rm z}^{\rm T}$ & $V_{\rm flat}$ \\
		  & [$10^{12}\Msun$] & [kpc] & [$10^{10}\Msun$] & [$10^{10}\Msun$] & [kpc] & [kpc] &  &  & [kpc] & [kpc] & [km/s] \\
		\hline
		g2.79e12 & 3.13 & 306 & 15.9 & 18.48 & 5.57 & 1.08 & 1.33 & 0.60 & 0.3 & 1.3 & 324 \\
		g1.12e12 & 1.28 & 234 & 6.32 & 7.93 & 5.69 & 0.77 & 1.53 & 0.04 & - & - & 166 \\
		g8.26e11 & 0.91 & 206 & 3.40 & 6.09 & 5.12 & 1.82 & 1.31 & 0.47 & 0.4 & 1.4 & 201 \\
        		g7.55e11 & 0.85 & 201 & 2.72 & 6.79 & 4.41 & 1.52 & 1.00 & 0.52 & 0.4 & 1.4 &181 \\
		g7.08e11 & 0.55 & 174 & 2.00 & 3.74 & 3.90 & 1.79 & 1.10 & 0.56 & - & 1.0 & 174 \\
		g6.96e11 & 0.68 & 187 & 1.58 & 4.79 & 5.70  & 2.96 & 1.39 & 0.41 & - & 1.4 & 165\\
        \hline
\end{tabular}
\end{center}
\end{table*}

\section{Cosmological zoom-in Simulations} \label{sec:simulation}

The six simulations analyzed in this work, g2.79e12, g1.12e12, g8.26e11, g7.55e11, g7.08e11 and g6.96e11 are higher-resolution versions of galaxies taken from the Numerical Investigation of a Hundred Astronomical Objects (NIHAO) simulation suite \citep{Wang2015} where the name designates the halo mass of the corresponding dark matter only run. A subsample of the galaxies presented here have already been used to study the build-up of the peanut-shaped bulge \citep{Buck2018,Buck2019b} or the dwarf galaxy inventory of MW mass galaxies \citep{Buck2019}. In a range of papers the NIHAO galaxies have been compared to observed properties of galaxies and shown to match the observed properties well, as for example the distribution of metals in the Circum Galactic Medium \citep{Gutcke2017}, Tully-Fisher relations \citep{Dutton2017}, and the structure of stellar and gaseous disks \citep{Obreja2016,Buck2017}. 

The galaxies are selected from the lower resolution NIHAO sample by mass only ($0.5\times10^{12}\lesssim M_{200}\lesssim3\times10^{12}\Msun$), without considering the stellar disk size, halo structure or merger history. For technical reasons with the zoom-in technique, we require the haloes to be isolated. Formally, we only consider haloes that have no other haloes with mass greater than one-fifth of the virial mass within $3$ virial radii.
The mass resolution of the simulations used in this work is a factor of 8 to 16 higher compared to the standard NIHAO resolution (see table \ref{tab:res}) such that it ranges between $m_{\rm dark}\sim1.5 - 5.1\times10^5 \Msun$ for dark matter particles and $m_{\rm gas}\sim2.8 - 9.4\times10^4 \Msun$ for the gas particles. The corresponding force softenings are $\epsilon_{\rm dark}=414 - 620$ pc for the dark matter particles and $\epsilon_{\rm gas}=177 - 265$ pc for the gas and star particles (see also table \ref{tab:res}). However, the adaptive smoothing length scheme implies that $h_{\rm  smooth}$ can be as small as $\sim50$ pc in the disk mid-plane. Figure \ref{fig:mass_res} shows a comparison of the resolution in terms of dark matter particles per halo between our simulations and a selection of both zoom-in and large box simulations in the literature. Our fiducial simulations are shown with colored points while the NIHAO sample is represented by the red open dots. Other zoom-in simulations in the literature \citep{Guedes2011,Governato2012,Marinacci2014,Sawala2016,Wetzel2016,Grand2017,Christensen2018,Hopkins2018,Garrison-Kimmel2018,Garrison2019} are shown with gray open symbols as indicated in the legend. Large box simulations \citep{Schaye2015,Dolag2016,Pillepich2018,Pillepich2019} are indicated with dashed/dotted lines.
Note, the simulations used in this work have a factor of 5 better resolution compared to e.g. TNG50 \citep{Pillepich2019} although slightly lower resolution than e.g. the Latte simulation \citep{Wetzel2016}. 

All NIHAO galaxies adopt cosmological parameters from the \cite{Planck}, namely: \OmegaM=0.3175,   \OmegaL=0.6825, \Omegab=0.049, H${_0}$  =  67.1\kms\Mpc$^{-1}$, \sig8 = 0.8344. The initial conditions are created the same way as for the original NIHAO runs \citep[see][]{Wang2015} using a modified version of the \texttt{GRAFIC2} package \citep{Bertschinger2001,Penzo2014}. The hydrodynamics, star formation recipe and feedback schemes exploited are the same as for the original NIHAO runs and are summarized below.

\subsection{Hydrodynamics}

The simulations were run with a modified version of the smoothed particle hydrodynamics (SPH) solver {\texttt{GASOLINE2}} \citep{Wadsley2017} with substantial updates to the hydrodynamics as described in \citep{Keller2014}. The modifications of the hydrodynamics improve multi-phase mixing and remove spurious numerical surface tension by calculating the ratio of pressure $P$ and density  $\rho$, $P/\rho^2$ as a geometrical average over the particles in the smoothing kernel as proposed  by \cite{Ritchie2001}. The treatment of artificial viscosity has  been modified to use the signal velocity as described  in \cite{Price2008} and the \cite{Saitoh2009} timestep limiter was implemented  so that cool particles behave correctly when a hot blastwave hits them. All simulations use the Wendland C2 smoothing kernel \citep{Dehnen2012} to avoid pairing instabilities and a number of 50 neighbour particles in the calculation of the smoothed hydrodynamic properties. All simulations in the NIHAO project, including the ones used here, employ a pressure floor to keep the Jeans mass of the gas resolved and suppress artificial fragmentation \citep[see also appendix A1 of][]{Smith2018}. Our implementation follows \citet{Agertz2009} which is equivalent to the criteria proposed in \citet{Richings2016} and fulfils the \citet{Truelove1997} criterion at all times. Thus, the Jeans mass in our simulations is resolved with $\sim 4$ SPH kernel masses.

Gas cooling in the temperature range from 10 to $10^9$ K is implemented via hydrogen, helium, and various metal-lines calculated using \texttt{cloudy} \citep[version 07.02;][]{Ferland1998} as described in \cite{Shen2010}. Cosmic reionization from the UV background of \cite{Haardt2005} is adopted. Finally, we adopted a metal diffusion algorithm between particles as described in \cite{Wadsley2008}.

\begin{figure*}
\begin{center}
\includegraphics[width=.75\textwidth]{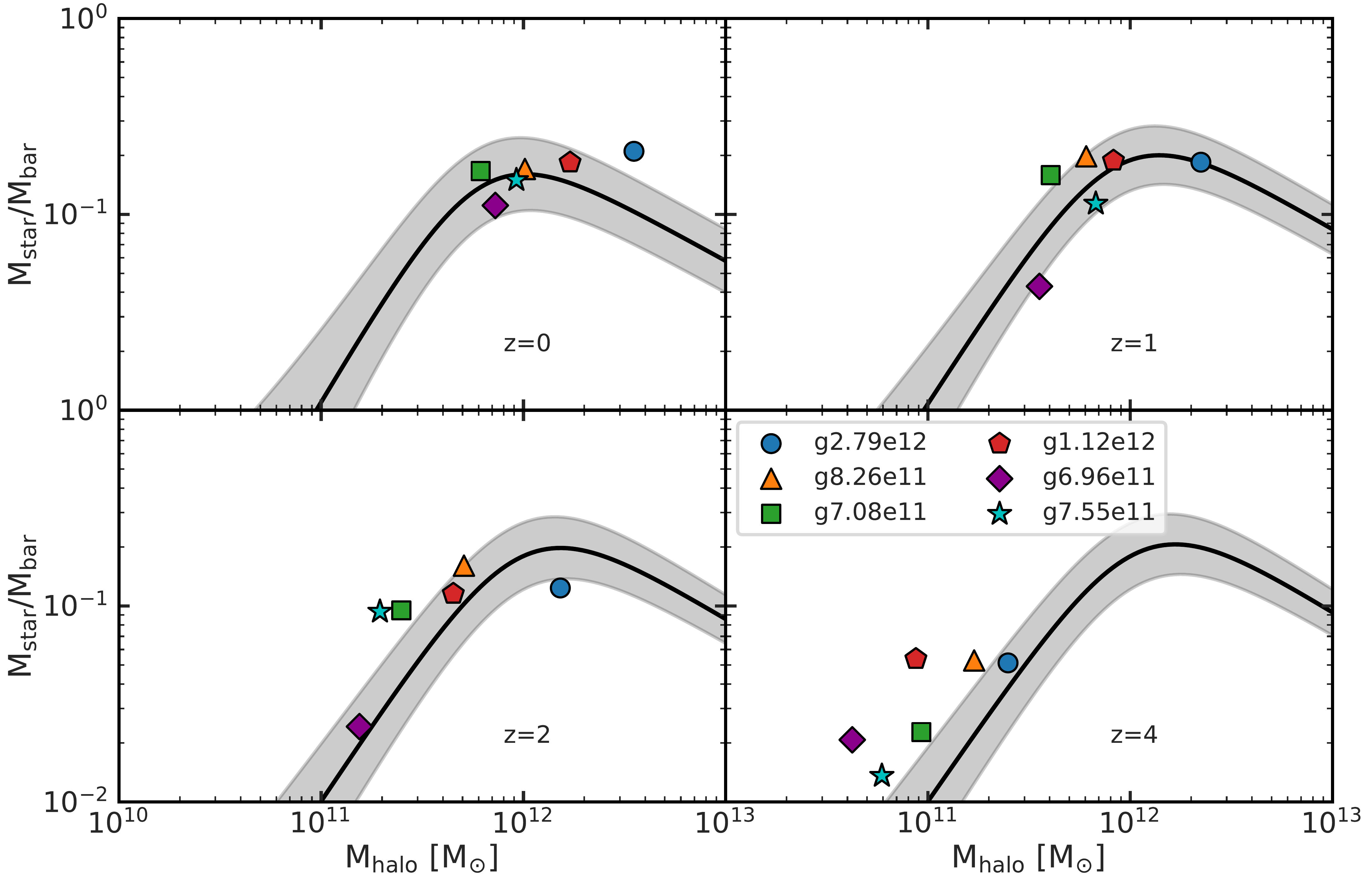}
\end{center}
\vspace{-.35cm}
\caption{The ratio between total stellar mass and total baryonic mass as a function of peak halo mass for 4 different redshifts as indicated in the panels. Colored symbols show the six galaxies in comparison to the relation from \citet{Moster2018} shown with the black line and the shaded gray area.}
\label{fig:baryon_conv}
\end{figure*}

\subsection{Star Formation and Feedback}
The  simulations employ the star formation recipe as described in \cite{Stinson2006}. Gas is eligible to form stars when it is dense ($n_{\rm  th}  >  10.3$cm$^{-3}$) and cold (T $< 15,000$K) such that the Kennicutt-Schmidt relation is reproduced. The threshold number density $n_{\rm th}$ of gas is set to n$_{\rm th} = 50 m_{\rm  gas}/\epsilon_{\rm gas}^3 = 10.3$ cm$^{-3}$, where $m_{\rm gas}$ denotes the gas particle mass and $\epsilon_{\rm gas}$ the gravitational softening of the gas and the value of 50 denotes the number of neighbouring particles. We fix the initial stellar particle mass at birth to $1/3\times m_{\rm{gas}}$ of the initial gas mass and apply mass loss according to stellar evolution models. A systematic study of the impact of the star formation threshold density on the galaxy properties has recently been conducted by \citet{Dutton2019}. \citet{Buck2019} showed that only simulations with a high ($n>10$) threshold density reproduces the observed spatial clustering of young star clusters.  

Two modes of stellar feedback are implemented as described in \cite{Stinson2013}. The  first  mode  models the energy  input from  stellar  winds and  photoionization from  luminous young  stars. This mode happens  before any supernovae explode and consists of the total stellar flux,  $2 \times  10^{50}$  erg of  thermal energy  per $M_{\odot}$ of the entire stellar population. The efficiency parameter for coupling the energy input is set to $\epsilon_{\rm ESF}=13\%$ \citep{Wang2015}.

The second mode models the energy input via supernovae type Ia and II. Feedback from supernovae type II starts 4 Myr after the formation of the star particle (lifetime of the most massive stars going off as supernovae type II) while feedback from supernovae type Ia are delayed according to the model of \citet{Raiteri1996}. Energetic feedback from supernovae is implemented using the blastwave formalism as described in \cite{Stinson2006} and applies a delayed cooling formalism for particles inside the blast region in order to compensate for numerical losses of unresolved supernova blastwaves. We refer the reader to \cite{Stinson2013} for further information and an extended feedback parameter search. 

\section{Galaxy properties}
\label{sec:G_props}

\begin{figure}
\begin{center}
\includegraphics[width=\columnwidth]{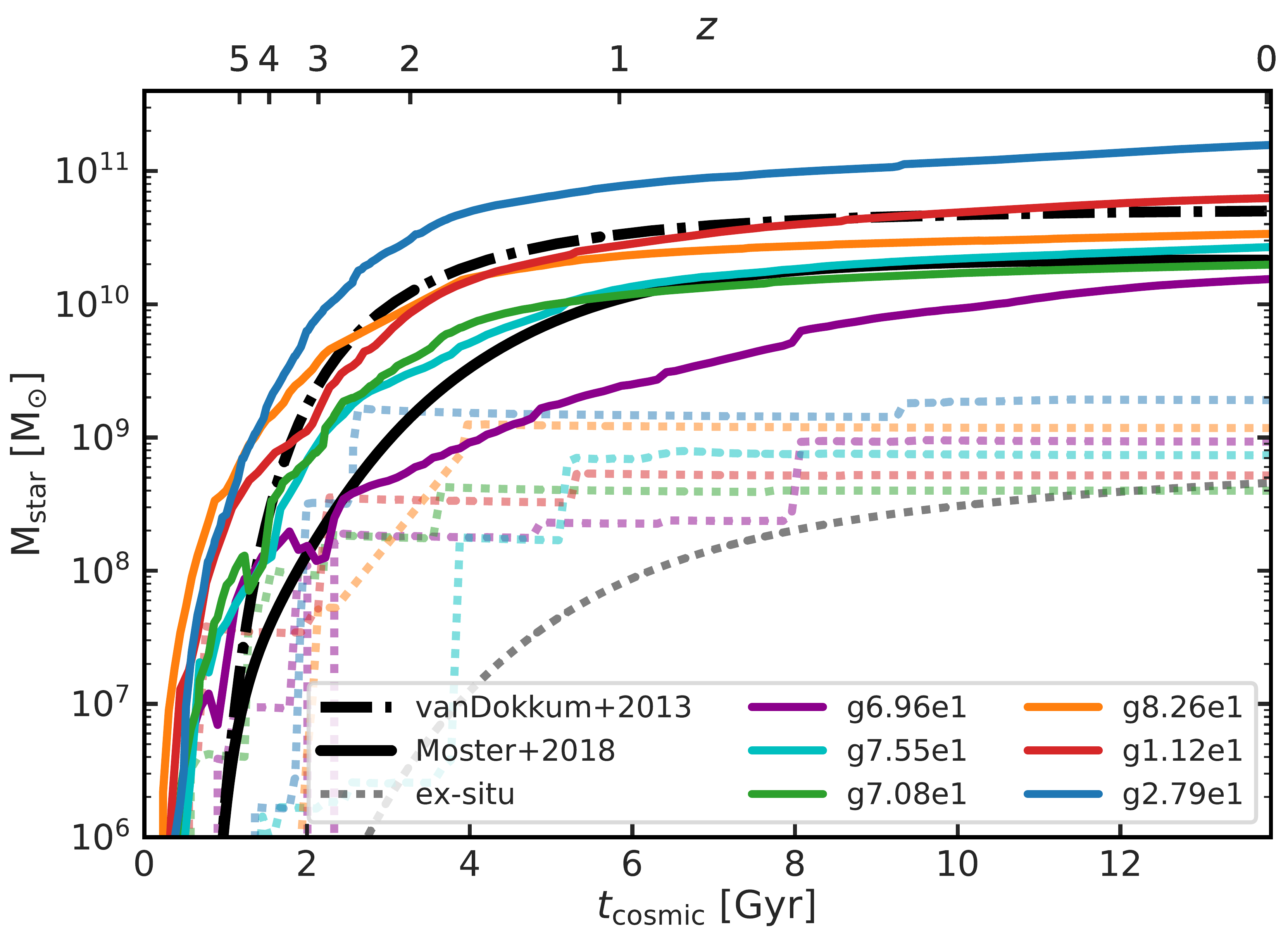}
\end{center}
\vspace{-.35cm}
\caption{Stellar mass growth vs. time. Colored solid lines show the increase of total stellar mass over time while colored dashed lines show the amount of accreted stellar mass. We compare the simulation results to observations of Milky Way progenitors from \citet{vanDokkum2013} (dashed black line) and predictions from \citet{Moster2018} for the total stellar mass (solid black line) and the accreted stellar mass (dotted black line).}
\label{fig:mass_growth}
\end{figure}

\begin{figure*}
\begin{center}
\includegraphics[width=\textwidth]{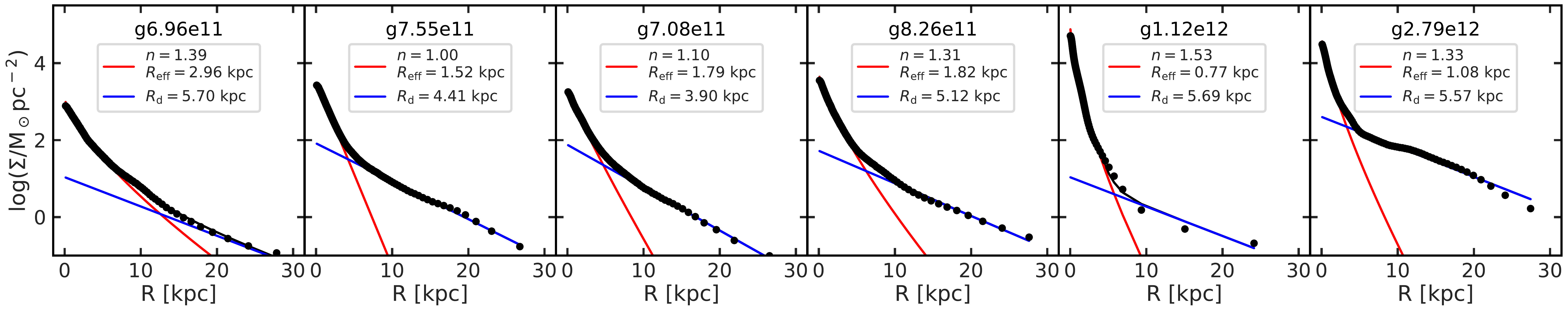}
\end{center}
\vspace{-.35cm}
\caption{Azimuthally averaged stellar mass surface density profiles. We simultaneously fit the profiles with a \citet{Sersic1963} (red line) and an exponential (blue line) profile where we fit the surface density out to $2 R_{90}$. The combined profile is shown with a black solid line. The resulting values for the S\'ersic index, the effective radius and the disk scale length are indicated in each panel.}
\label{fig:surf_den}
\end{figure*}

For this work the virial mass, $M_{200}$, of each isolated halo is defined as the mass of all particles within a sphere containing $\Delta$ = 200 times the cosmic critical matter density, $\rhocrit$. The virial radius, $R_{200}$, is defined accordingly as the radius of this sphere. The haloes in the zoom-in simulations were identified using the MPI+OpenMP hybrid halo finder \texttt{AHF2} \citep{Gill2004,Knollmann2009}. Merger trees are calculated using the analysis tool \texttt{tangos} \citep{tangos} by tracing the particle IDs of all dark matter particles through time and identifying all progenitor halos of a given galaxy/dark matter halo at redshift $z=0$. Detailed parameters for each galaxy including the total virial mass, virial radius, gas and stellar mass or rotation velocity can be found in table \ref{tab:props}.

The six galaxies used for this work span the halo mass range of $6.8\times10^{11}\Msun$ to $3.1\times10^{12}\Msun$ and have stellar masses between $3.4\times10^{10}\Msun$ and $1.59\times10^{11}\Msun$. Figure \ref{fig:impression} shows an overview of the main galaxies formed in the simulation. From left to right we show the dark matter and stellar surface density with the dashed white circle indicating the virial radius. We further show a closer in view of the HI disk and an RGB rendering of the stars in the inner $75$ kpc. Except for galaxy g1.12e12 all simulations result in thin disk galaxies of scale length $\sim5$ kpc and total scale heights of $\lesssim1$ kpc as can be seen from the right most panels of figure \ref{fig:impression}. The redshift zero snapshots and halo catalogues of the six galaxies are publicly available for download here: \url{{http://www2.mpia-hd.mpg.de/~buck/#sim_data}}.
Access to higher redshift outputs may be granted upon email contact.

In what follows, we study in detail the properties of these galaxies such as stellar masses and sizes, rotation velocities or structure of the stellar disk and compare them to observed properties of MW analogues. In a forthcoming paper we will further study in detail the stellar chemical abundance space of these galaxies (Buck et al. in prep.).

\subsection{Stellar mass growth}

One particularly strong test for the employed feedback model of the simulations is the baryon conversion efficiency $M_{\rm star}/M_{\rm bar}$, the ratio of stellar mass to total baryonic mass. A value of one would mean the galaxy turned all its available cosmic baryon budget into stars and indicates a failure of the employed feedback. Empirically the baryon conversion efficiency is constrained to be a strong function of halo mass \citep[e.g.][]{Guo2010,Behroozi2013,Moster2013,Kravtsov2018,Moster2018,Behroozi2019} with a peak efficiency of $\sim20\%$ at halo masses of $\sim10^{12}\Msun$ and strong decline at lower and higher masses. 

In figure \ref{fig:baryon_conv} we compare the baryon conversion efficiency $M_{\rm star}/M_{\rm bar}$ of the simulated MW analogues (colored points) to empirical estimates from \citet[][black line]{Moster2018} for four different redshifts. The simulations follow the relation from \citet{Moster2018} very well at redshifts below $z=4$. The only strong deviation being g2.79e12 at $z=0$ which is caused by a too massive bulge owing to the missing super-massive black hole feedback. The discrepancy can already be alleviated by inclusion of local photoionization feedback as shown in \citet{Obreja2019}.
At redshift $z=4$ our simulation show baryon conversions slightly higher than predicted by the empirical model. However, this is no major issue as the stellar masses of the simulated galaxies at these high redshifts agree well with the observed stellar masses of MW progenitors.

\citet{vanDokkum2013} constructed the stellar mass growth of MW analogue galaxies from the 3D-HST and CANDELS Treasury survey \citep{3D-HST,CANDELS}. In figure \ref{fig:mass_growth} we compare the results from \citet[][black dashed line]{vanDokkum2013} to the stellar mass growth in our simulations (colored solid lines). Additionally, we also add the estimates from \citet{Moster2018} for the stellar mass growth of star forming galaxies with the solid black line. We find very good agreement between the stellar mass growth of our simulations and the inferred stellar mass growth for MW analogues from \citet{vanDokkum2013} with a strong increase in stellar mass in the early universe up to an age of $\sim4$ Gyr ($z \lesssim 2$) followed by a strong flattening of the growth rate. A similar but somewhat slower growth in stellar mass is predicted by the empirical model of \citet{Moster2018}. This discrepancy between the \citet{Moster2018} results and the \citet{vanDokkum2013} results reinforces the fact that `out missing' the baryon conversion relation at $z=4$ is not a real issue here. 

We further highlight in this figure the amount of stellar mass contributed to the main galaxy by merging satellites. With dotted lines we show the amount of ex-situ stars in the simulations (colored lines) and predicted by the empirical model of \citet{Moster2018}. In any case, at redshift $z=0$ we find a fraction of at most a few percent of ex-situ stars mostly brought into the galaxy at early times before redshift $2-1$ in good agreement with \citet[][dotted black line]{Moster2018} and results from the AURIGA simulations \citep{Grand2017,Monachesi2019} but somewhat smaller than previous findings \citep{Pillepich2015}. This result has strong implications on the presence and possibilities of finding such ex-situ stars in the MW brought in by disrupted satellite galaxies \citep[e.g.][]{Belokurov2018,Koppelman2018,Helmi2018} and further analysis is needed to help understand the relative importance of accreted stars vs in-situ stars. Some studies have proposed that disk stars can be heated by satellite galaxies into the stellar halo of the MW \citep[e.g.][]{Haywood2018} although a recent study of the MW's stellar halo using Gaia DR2 data by \citet{Deason2019} shows that the relatively low stellar halo mass of the MW can only be reconciled if the MW did not suffer any significant late time ($t\lesssim10$ Gyr) merger. In the remainder of this paper we study the structure of the stellar disks in detail. We note that the galaxy g1.12e12 is a pure spheroidal galaxy without any significant stellar disk. This galaxy undergoes a major merger early in its history ($z\sim2.5$), after which the stellar disk is destroyed and a compact morphology is established and until redshift $z=0$ no stellar disk is reformed. The current simulation setup does not include AGN feedback, which might explain the compactness of this galaxy. We plan to re-run some of these simulations, including g1.12e12, with the Gasoline2 code version including the AGN feedback prescriptions by \citet{Blank2019} with the goal of better understanding the effect of AGN on galaxy morphology. Here we drop this galaxy from our sample from now on.

\begin{figure*}
\begin{center}
\includegraphics[width=.49\textwidth]{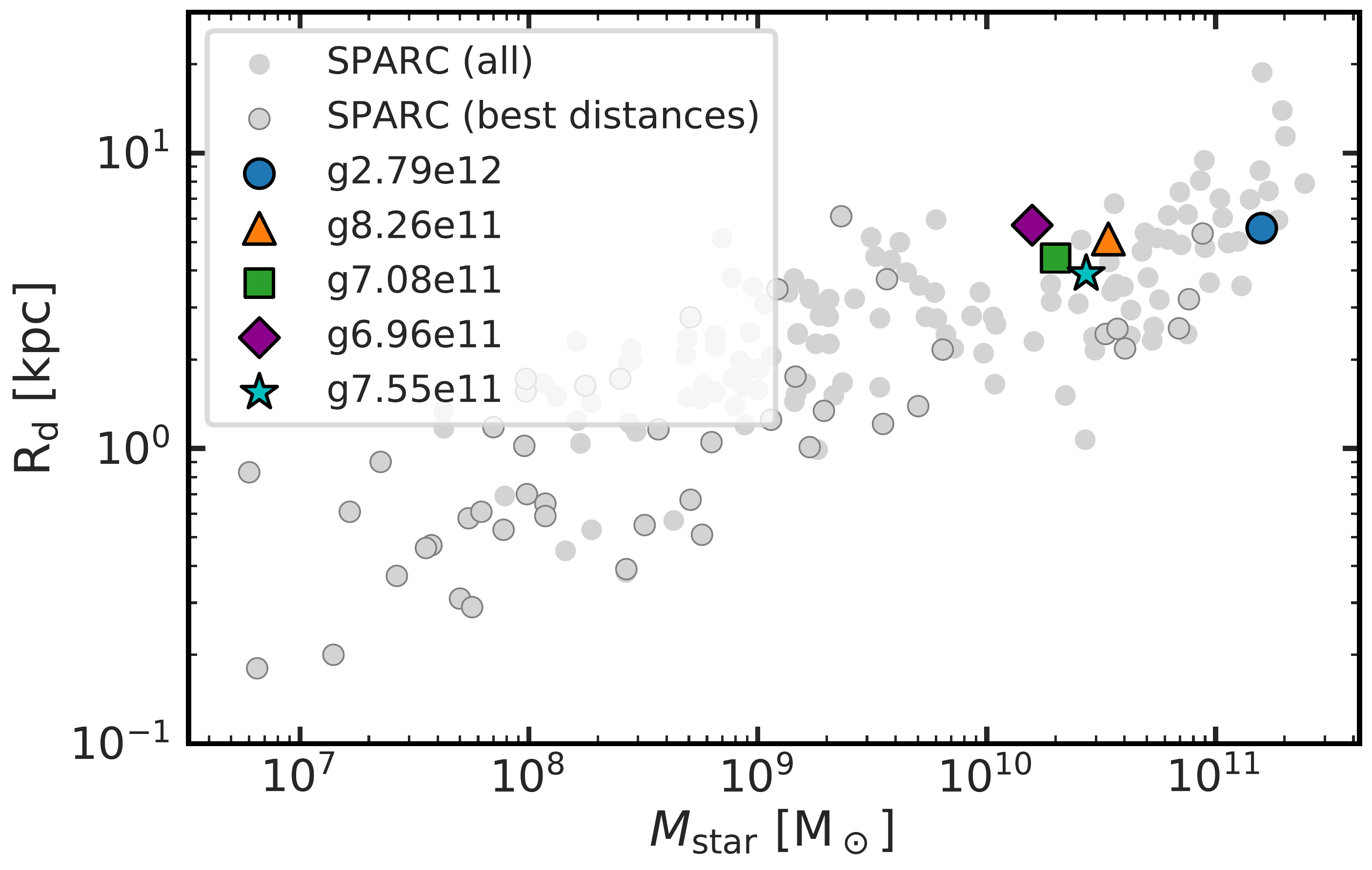}
\includegraphics[width=.5\textwidth]{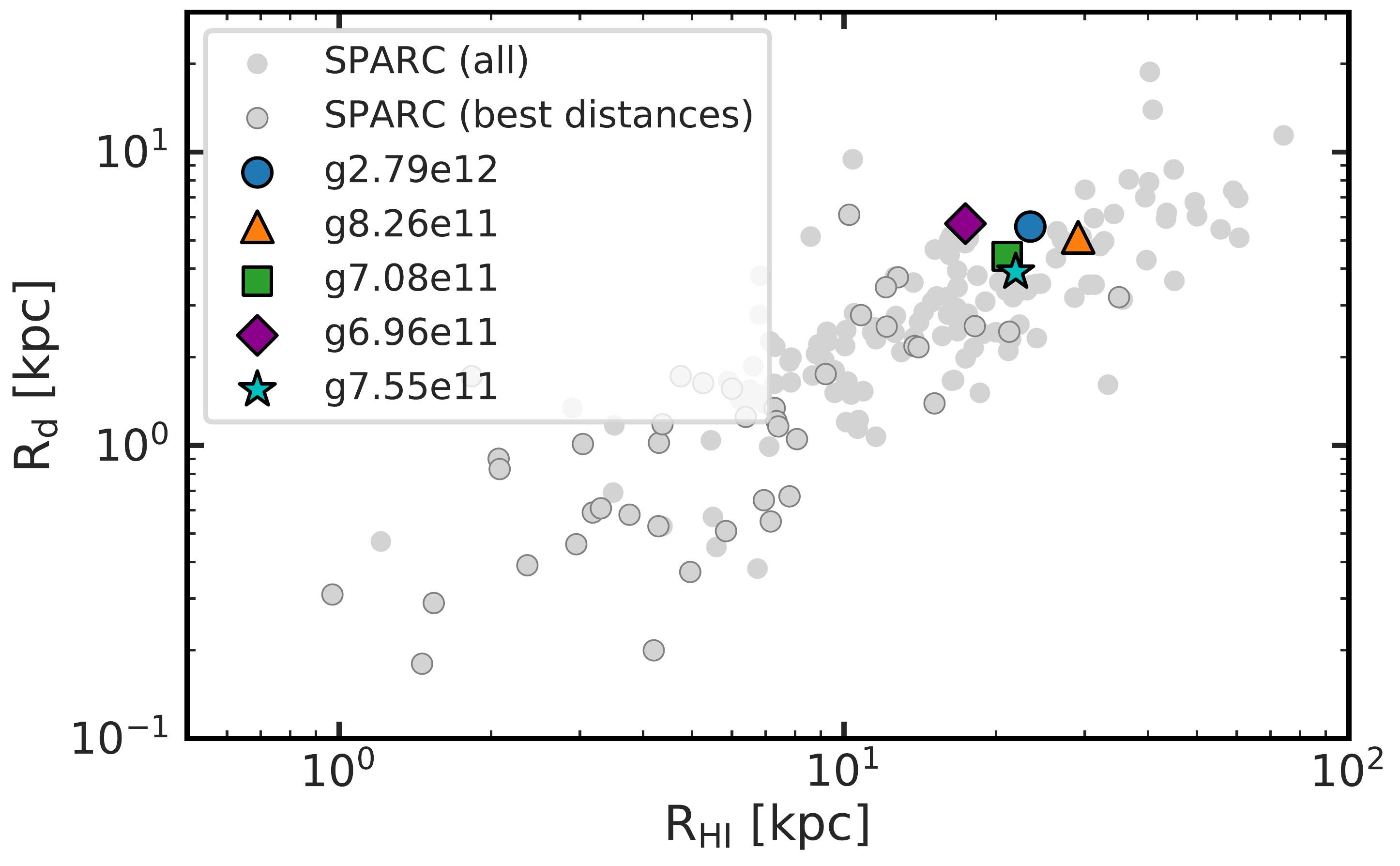}
\includegraphics[width=.49\textwidth]{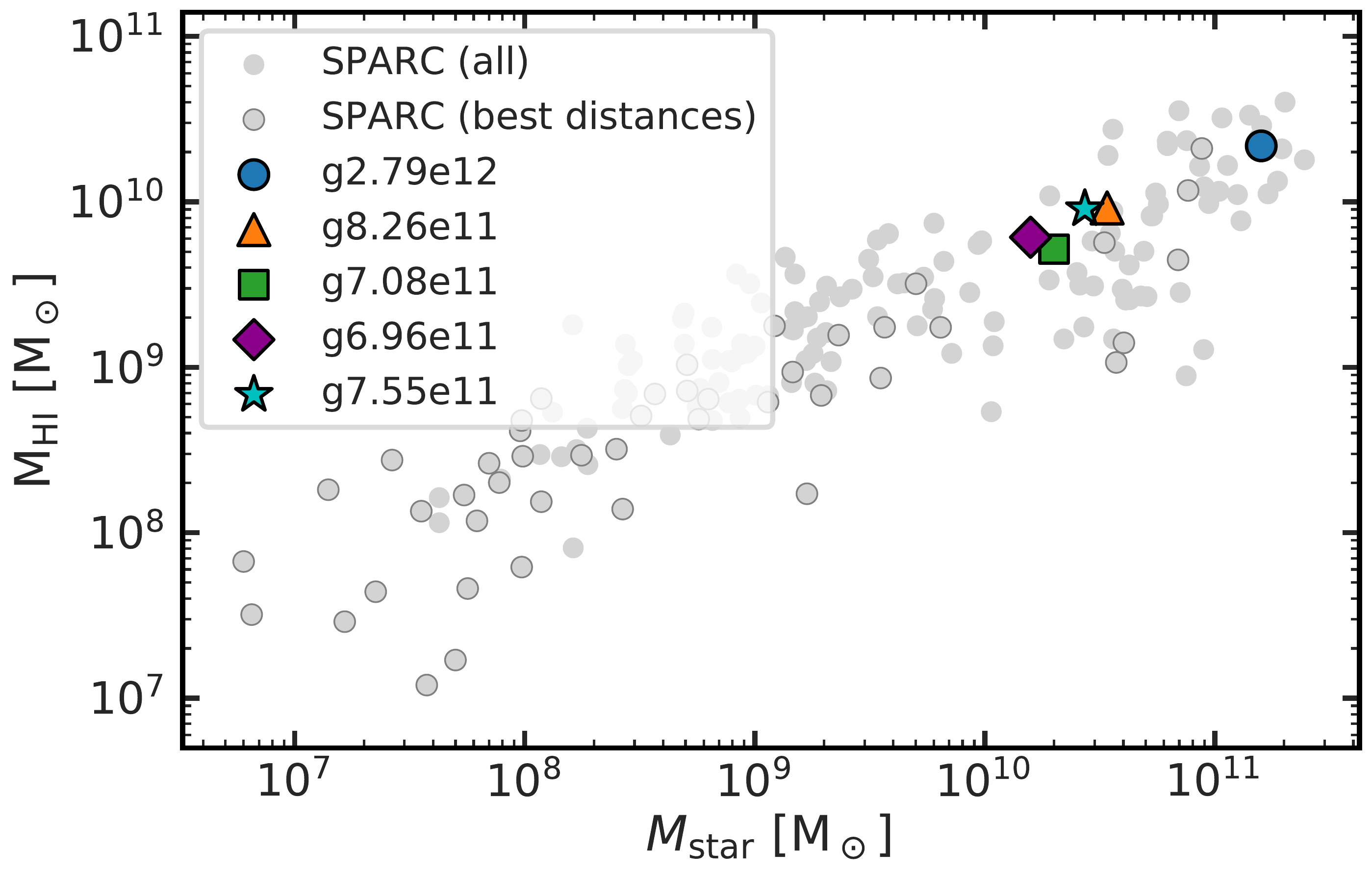}
\includegraphics[width=.5\textwidth]{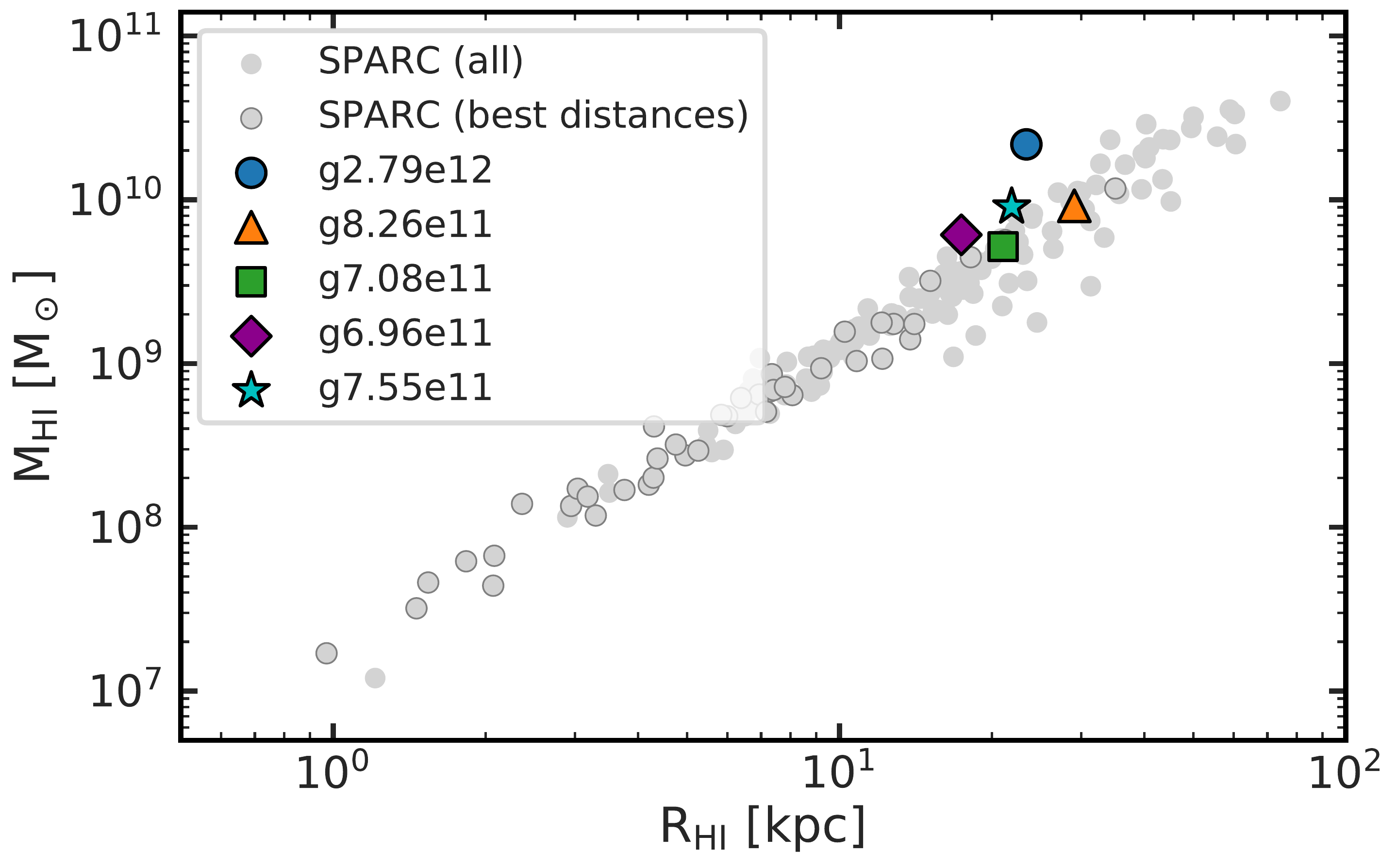}
\end{center}
\vspace{-.35cm}
\caption{Comparison with SPARC observations: \emph{Upper left:} disk scale length vs. stellar mass. \emph{Upper right:} disk scale length vs. HI disk size $R_{\rm HI}$. \emph{Lower left:} HI mass $M_{\rm HI}$ vs. stellar mass. \emph{Lower right:} HI mass $M_{\rm HI}$ vs. HI disk size $R_{\rm HI}$. The fainter gray symbols represent the complete SPARC sample, while the ones circled by a black border highlight the galaxies with very good quality distance data.}
\label{fig:sparc}
\end{figure*}

\subsection{Stellar disk sizes}

Reproducing the correct amount of stellar mass is one test for the employed feedback model, however even more important is to reproduce the spatial structure of the stellar component. In order to compare the sizes of our simulations to observations of both the MW and local spiral galaxies from the SPARC survey \citep{Lelli2016} we fit the stellar surface density simultaneously with a \citet{Sersic1963} plus an exponential profile. We show the results of this exercise in figure \ref{fig:surf_den} where dots show the data, black solid lines the combined fit, red lines the S\'ersic component and blue lines the exponential part. Resulting values for the fit parameters are indicated in each panel. Disk scale length range in between $\sim4-5$ kpc  in good agreement with observations of local disk galaxies \citep{Lange2015} but somewhat larger than the MW's scale length with $R_{\rm s}^{\rm MW}\sim2.6\pm0.5$ kpc \citep[see][for details]{Bland-Hawthorn2016} with the highest value of $3.9$ kpc measured by the GLIMPSE survey \citep{Benjamin2005}. The effective radii for the bulges are of the order of $\sim1.0-1.3$ kpc and S\'ersic indices range between $n\sim1-1.5$. The surface density profiles of our galaxies are reasonably stable across resolution levels. In the appendix, in figure \ref{fig:surf_res_comp} we compare the surface density profiles of the lower resolution versions to our high resolution galaxies finding that within $R_{90}$ both profiles agree well.

\begin{figure*}
\begin{center}
\includegraphics[width=\textwidth]{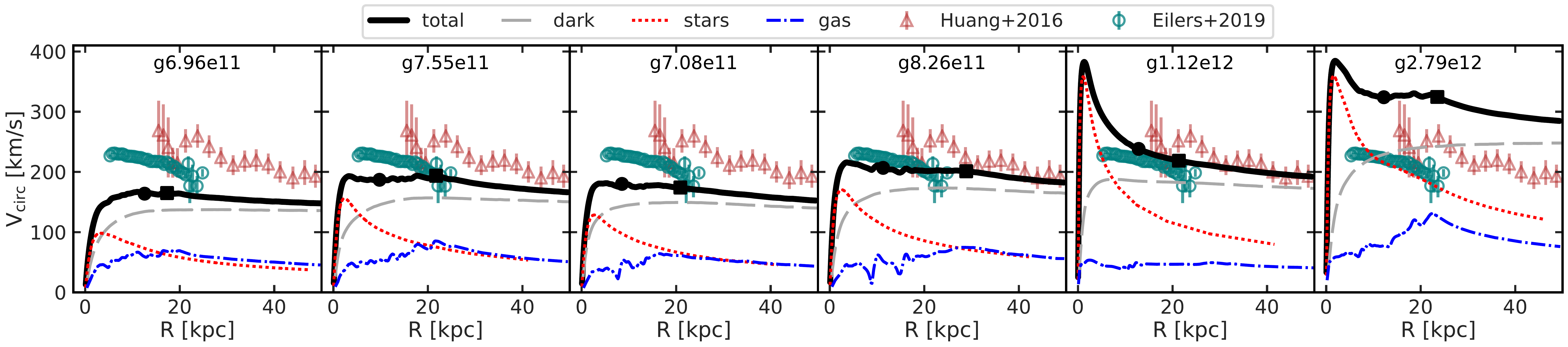}
\end{center}
\vspace{-.35cm}
\caption{Circular velocity curves $v=\sqrt{GM(R)/R}$ for all matter components (black), only stars (red dotted), gas (blue dashed) and dark matter (gray long-dashed). The black dot indicates the radius at which $2.2R_{\rm d}$ is reached and the black square indicates the HI disk radius where we measure the circular velocity of the flat part of the rotation curve. Observational data for the MW from \citet{Eilers2019} and \citet{Huang2016} are shown with green dots and red triangles respectively. The galaxies simulated for this work span a range in stellar masses bracketing the observed rotational velocity of the MW with g8.26e11 being the best MW analogue.}
\label{fig:rot}
\end{figure*}
 
\begin{figure*}
\begin{center}
\includegraphics[width=\columnwidth]{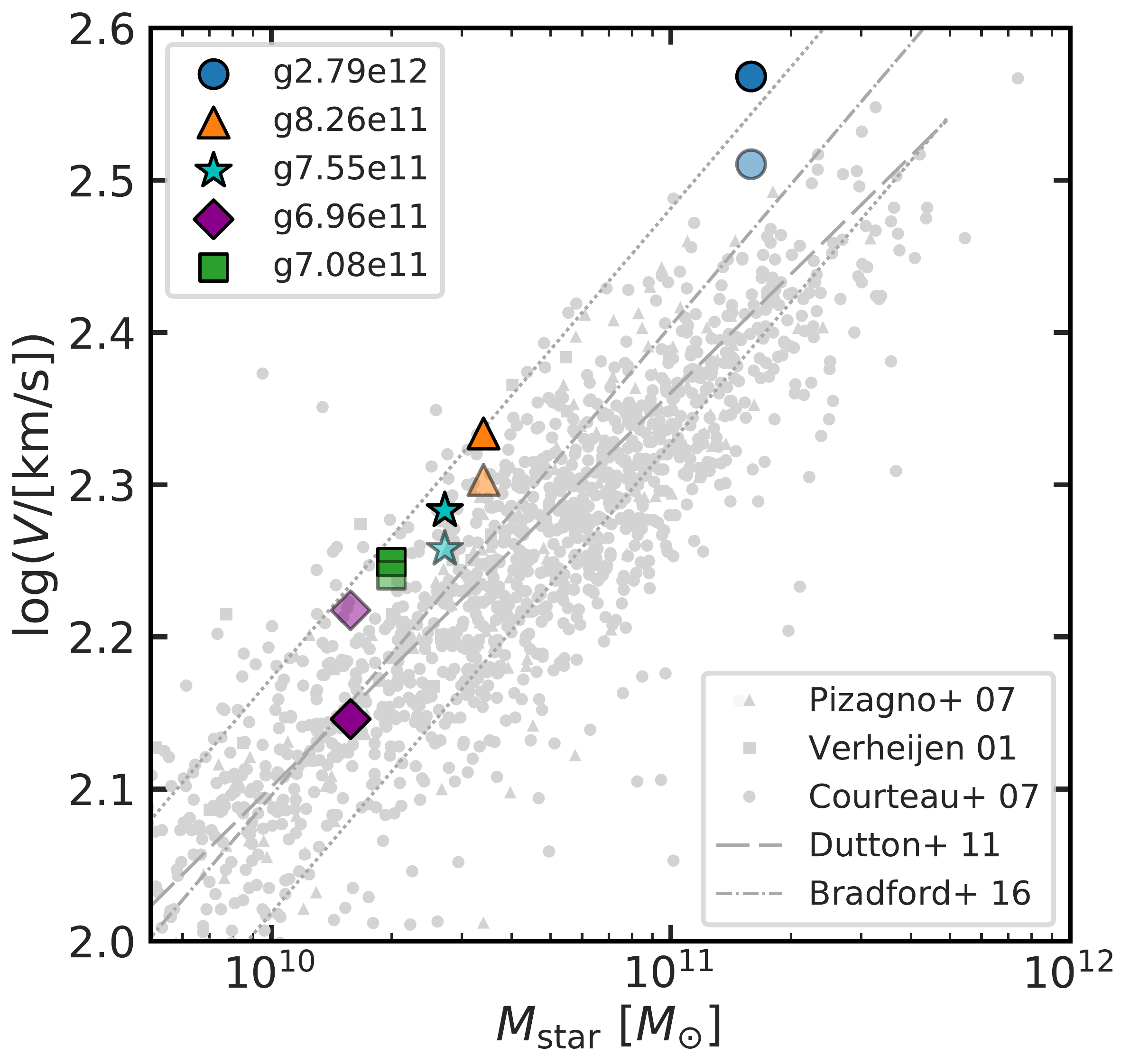}
\includegraphics[width=\columnwidth]{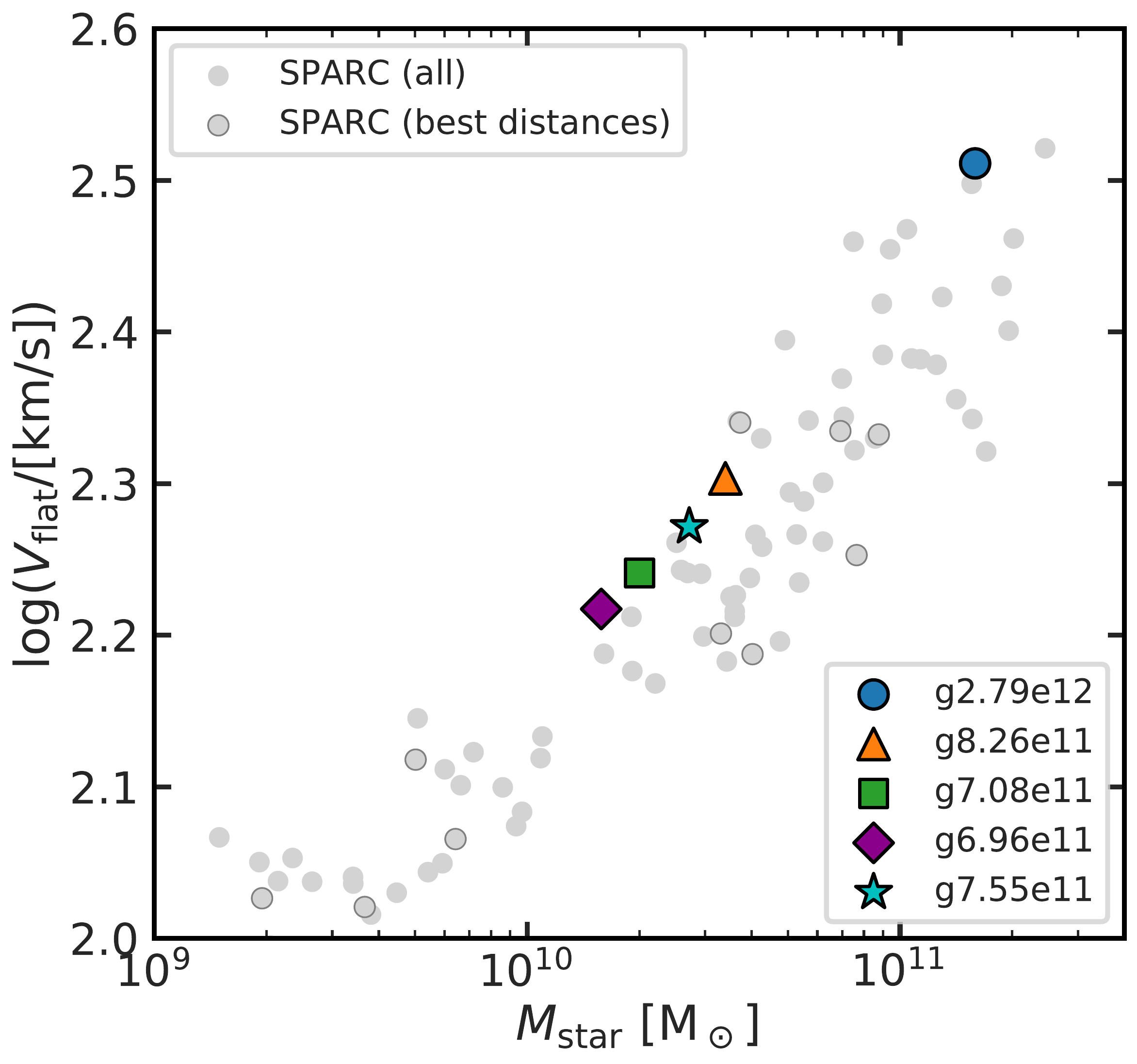}
\end{center}
\vspace{-.35cm}
\caption{\emph{Left panel:} Tully-Fisher relation for the simulations in comparison to observational data from \citet[][]{Verheijen2001,Pizagno2007,Courteau2007} shown with gray dots and \citet[gray dashed line]{Dutton2011}. Solid coloured points show the circular velocity measured at $2.2R_{\rm d}$ while shaded coloured points show the rotation velocity measured via the edge-on HI line widths. \emph{Right panel:} Rotation velocity measured at the HI disk radius vs. stellar mass in comparison to the SPARC dataset.}
\label{fig:tully}
\end{figure*}

Next, in Fig. \ref{fig:sparc}, we compare the extracted disk scale length, the sizes and masses of the HI disks in our simulations to results from the SPARC sample. The two upper panels compare the disk scale length of simulations and observations in terms of stellar mass (left panel) and HI disk size (right panel) where the HI disk size is defined as the radius where the surface density of the HI gas disk drops below $1\Msun/$pc$^2$. 

We find that the sizes of our galaxies are in good agreement with the observations from the SPARC data. The two lower panels compare the HI masses as a function of stellar mass (left panel) and HI disk size (right panel). Also in this parameter space we find good agreement between observations and simulations, although the lower right panel leaves the impression that the NIHAO-UHD galaxies show HI masses on the higher end of the observed values. Only the galaxy g2.79e12 shows a slightly too compact HI disk given its mass. We checked that the results regarding the HI properties are robust against changes in the calculation of the HI fraction. For example, using the HI calculations presented by \citet{Rahmati2013} gives similar results as using the direct output from the simulation.

The gas mass of the MW is estimated to be around $\sim1.7\times10^{10}\Msun$ \citep[table 5][]{Bland-Hawthorn2016}. By comparison, our simulations have higher gas masses ($4\times10^{10} - 18\times10^{10} \Msun$). On the other hand, we have seen that the HI masses are in agreement with results from the SPARC sample. Further analysis of the gaseous properties as a possible probe of the simulation recipe is needed in order to draw strong conclusions. A first step in this direction was recently taken by \citet{Obreja2019} who have shown that the inclusion of local photoionization feedback reduces the gas mass by roughly a factor of $2$ \citep[e.g. fig. 3][]{Obreja2019} while only weakly effecting the stellar masses of MW-like galaxies. Thus, the gas masses are a sensitive probe of the physics included in the simulations and we expect that inclusion of photoionization feedback would bring the gas masses of our galaxies in better agreement with the MW.

\begin{figure*}
\begin{center}
\includegraphics[width=\columnwidth]{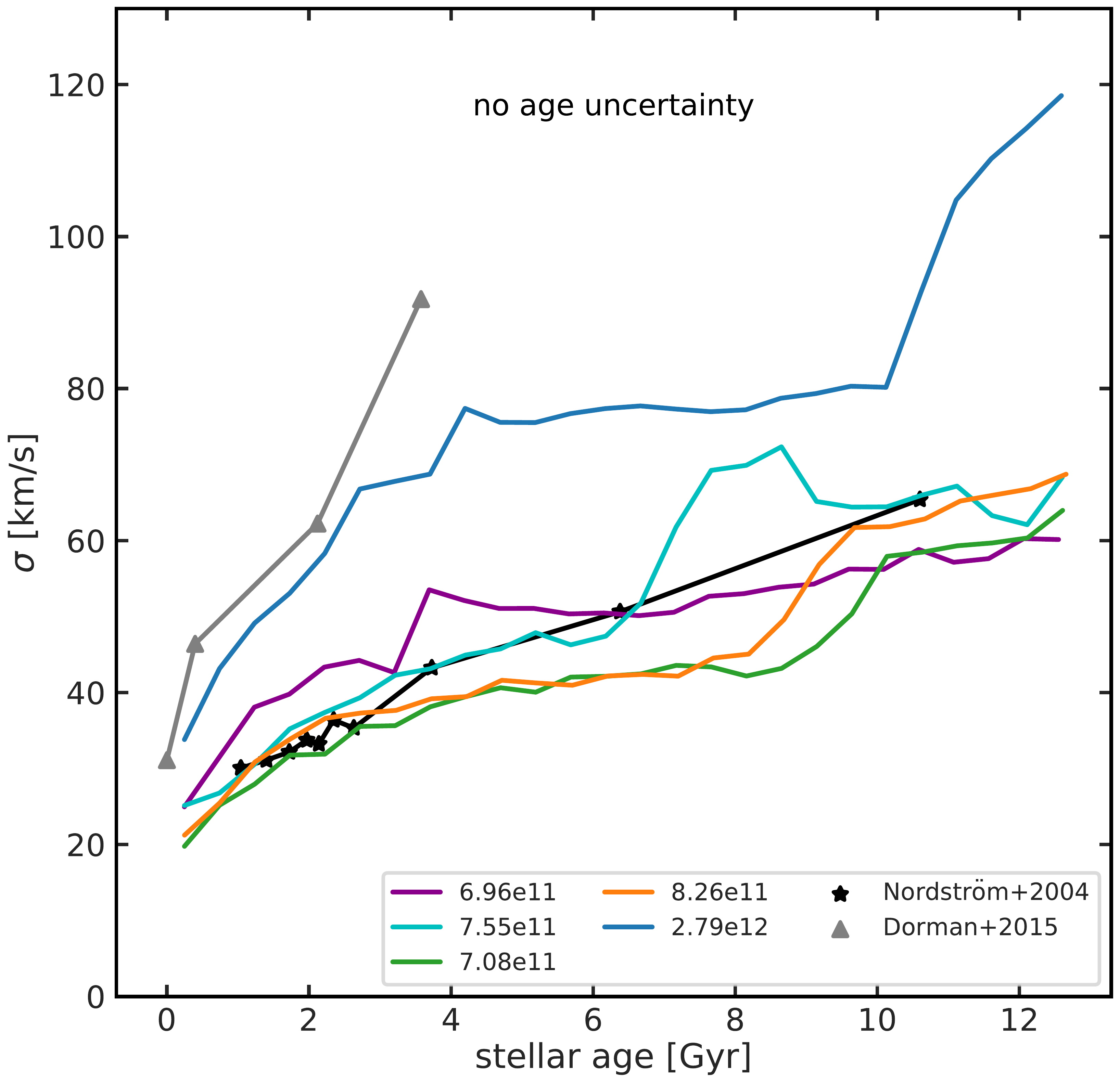}
\includegraphics[width=\columnwidth]{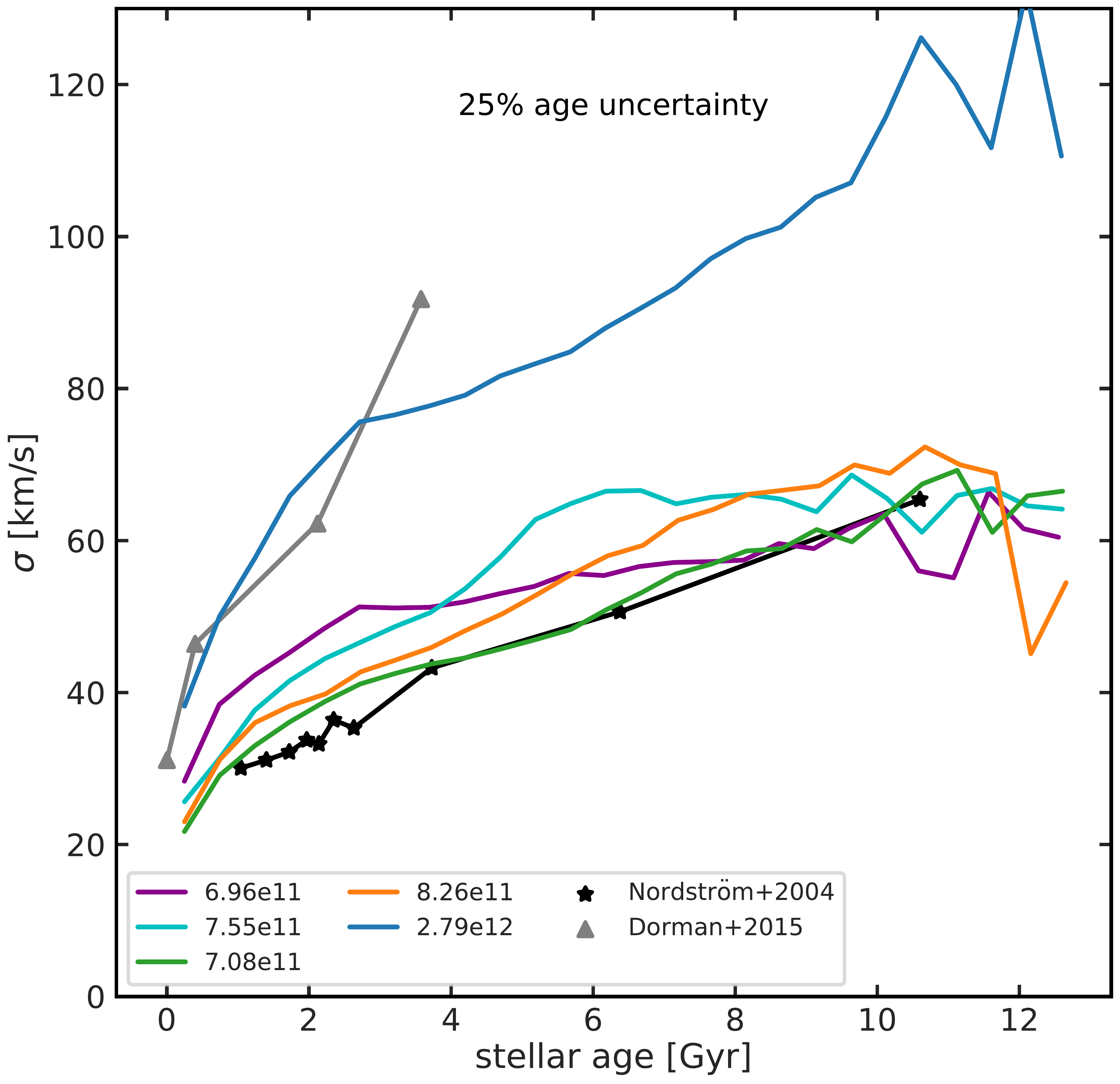}
\end{center}
\vspace{-.35cm}
\caption{Age velocity dispersion relation (AVR) for simulated disk stars (colored lines) in comparison to observational data obtained for the MW \citet[black stars]{Nordström2004} and M31 \citet[gray triangles]{Dorman2015}. The left panel shows the AVR without considering age uncertainty in the simulations while the right panel assumes $25\%$ age uncertainty for simulated disk stars. We only select stars within a radial annulus of width $2$ kpc around the solar radius $R_{0}=8$ kpc and a vertical height of $z<\vert2.5\vert$ kpc from the simulations. The quick rise in velocity dispersion at the oldest stars in the left panel is observed in the MW \citep[e.g.][]{Minchev2018} and has been suggested to result from dynamical heating due to the last massive merger \citep[e.g.][]{Martig2014b}. Note that this rise is completely erased when age uncertainties are included.}
\label{fig:vel_disp}
\end{figure*}

\subsection{Rotation curves}

Having established the realism of the simulated galaxies in terms of baryonic mass and size (both stellar and gaseous) we now turn to analyse their kinematic properties. In figure \ref{fig:rot} we show the circular velocity curve $v=\sqrt{GM(R)/R}$ for dark matter (gray dashed), gas (blue dashed dotted), stars (red dotted) and the total mass budget (thick black line). In order to guide the eye we also add observational data for the MW from \citet{Huang2016} and \citet{Eilers2019}. The simulations result in disk galaxies with exceptionally flat rotation curves with rotation velocities of $\sim200$ km/s except for the most massive galaxy g2.79e12 which shows a peaked rotation curve in the bulge region. The reason for this is the slightly too compact bulge. We can further appreciate that all galaxies are stellar mass dominated in their central regions ($R<5$ kpc) and outside this radius dark matter takes over. Gas and stars are of equal importance at radii of $\sim20$ kpc. 

We would like to highlight that the galaxy g8.26e11 is an ideal MW analogue in terms of its rotation curve when compared to the data of \citet{Eilers2019}, see also \citet{Obreja2018} for the NIHAO equivalent of this galaxy. However, a comparison to the dynamics of a larger sample of disk galaxies is needed in order to establish the realism of these simulations. Therefore, in the left panel of figure \ref{fig:tully} we show the stellar Tully-Fisher relation in comparison to the simulations (colored dots). Here the solid coloured dots are taken to be the rotation velocity at $2.2R_{\rm d}$ while the faint symbols represent the edge-on rotation velocity as derived from the HI line width. In the right panel we compare the circular velocity of the flat part of the rotation curve of the SPARC galaxies to the simulations where we measure the flat rotation velocity at the HI disk radius. Rotation curves of different resolution levels agree to better than $\sim10\%$ as we show in figure \ref{fig:rot_curve_comp} meaning that both the total mass and the structure are are reasonably stable between resolution levels.

The dynamics of the simulated galaxies are in broad agreement with the observations, both in terms of the stellar Tully-Fisher relation as well as in comparison to the SPARC data. However, we acknowledge the fact that in both cases the rotation velocity of the simulations resides on the high side of the observed values. We attribute this to the fact that the effective radii of the bulges are on the lower side of observed values (compare upper mid panel of fig. \ref{fig:sparc}). Especially the very good agreement of the galaxy g6.96e11 in both effective radii and rotation velocity is supporting this conclusion.

\subsection{Age velocity dispersion relation}

\begin{figure*}
\begin{center}
\includegraphics[width=\textwidth]{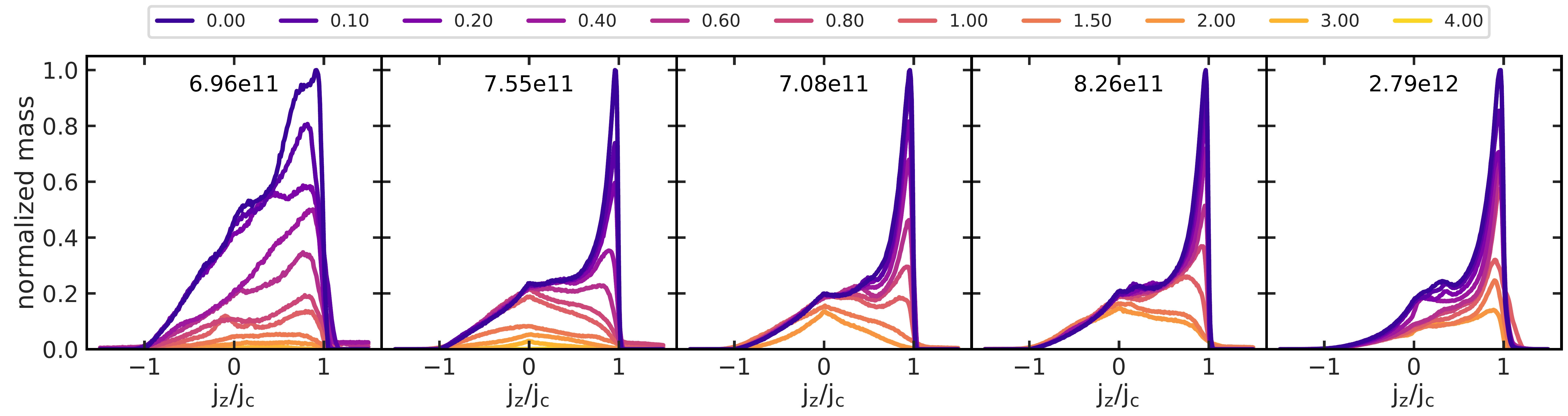}
\end{center}
\vspace{-.35cm}
\caption{Circularity distribution of stellar particles in the simulation for different redshifts as indicated by the line color. The stellar disk in the simulations is consistently build-up after redshift $\sim1$. The galaxy g6.96e11 show less ordered motions due to continued mergers impacting the stellar disk at lower redshifts.}
\label{fig:circ}
\end{figure*}

Finally, we compare the vertical velocity dispersion as a function of stellar age to observations of the MW. It has been well established that the velocity dispersion of a disk stellar population increases with age \citep[e.g.][]{Strömberg1925,Wielen1977}. 
In figure \ref{fig:vel_disp} we show the vertical velocity dispersion of star particles in the solar neighbourhood defined as an annulus of radius $R=8$ kpc of $2$ kpc thickness and vertical height of $z<\vert2.5\vert$ kpc as a function of stellar age. We compare the five disk simulations to observational data for the MW taken from \citep{Nordström2004} and M31 \citep{Dorman2015}. The upper panel assumes no age uncertainty for the stars in the simulation while the lower panel adopts $25\%$ age uncertainty. First of all, this figure shows that the age-velocity dispersion relation (AVR) in the simulations and the observations are in good agreement both for the MW data and M31's data. We would like to especially highlight the steep rise in velocity dispersion for old stellar ages $>8$ Gyr for most of the galaxies and the jump around $3$ Gyr for the galaxy g6.96e11 which is clearly seen in the upper panel. \citet{Martig2014b} found that such a jump might be caused by merger events which we confirm here. In particular, their simulations show that the slow rise in velocity dispersion is caused by subsequent heating of the stars rather than being set at birth. Interestingly, such a step in the AVR has been previously found observationally \citep{Quillen2001,Freeman2002} although already a $25\%$ age uncertainty in the simulations washes out the signal completely. A jump in $\sigma_{\rm vz}$ can also be seen in \citep{Minchev2018} using HARPS GTO data and appears much better for stars born in the inner disk, which is consistent with the effect of a last massive merger about $10$ Gyr ago.

These findings together with all other results presented in this section show that the employed feedback model of the NIHAO galaxies is able to model galaxies reproducing key observations of local disk galaxies including MW and M31. With this in mind we now turn to analyse the vertical and radial structure of our stellar disks in detail with the aim of understanding its build-up and evolution.

\section{Stellar disk structure}

It is still a matter of debate if simulations are able to model stellar disks of realistically thin scale heights. For the MW it is known that young stars reside in a stellar disk with a scale height $\sim220-450$ pc while the old stars in the thicker disk show scale heights of $\sim1450$ pc \citet{Bland-Hawthorn2016}. Furthermore, there is an ongoing discussion in the literature as to whether the stellar disks of galaxies flare, i.e. their scale heights increase with radius \citep[see discussion in][]{Minchev2015}. In this section we study in detail the structure of the stellar disks starting by analysing when and how the stellar disk is assembled.

\subsection{Angular momentum distribution}

In figure \ref{fig:circ} we show the distributions of stellar circularities, $j_{\rm z}/j_{\rm c}$, the ratio of stellar specific angular momentum and the angular momentum a circular orbit with the same energy would have \citep{Abadi2003,Brook2004,Tissera2012}. A value of $1$ corresponds to a perfect circular orbit and is usually associated with stars in the stellar disk while stars with smaller values have increasingly larger velocity components perpendicular to the stellar disk. These stars are usually associated with a spheroidal component. In figure \ref{fig:circ} we show the evolution of the stellar orbit circularities for several redshifts ranging from $z=4$ to $z=0$. Calculations of the stellar circularities have been performed using the software package \textit{galactic structure finder} \texttt{gsf} \citep{gsf,Obreja2018}\footnote{We calculate the gravitational potential from the simulations and smoothly interpolate it to calculate the values of $j_{\rm z}/j_{\rm c}$ for each star particles. If there are prominent substructures in the galaxies (e.g. at high redshift or stellar bars at low redshift) the smooth potential might slightly deviate from the true potential and values for $j_{\rm c}$ might be over/under estimated. Therefore we find a tiny fraction of star particles with $j_{\rm z}/j_{\rm c}$ values slightly above or below $1$.}. Consistent with previous findings we find that five of our galaxies are strongly disk dominated systems with most stars having circularities of $j_{\rm z}/j_{\rm c}>0.75$. We further see that the smallest galaxy g6.96e11 has a pretty large bulge component as can already be appreciated from the edge-on picture of this galaxy in figure \ref{fig:impression}.

Looking at the time evolution and thus the build-up of the stellar components we consistently find that in the early universe up to redshift $z\gsim1$ these galaxies were bulge dominated systems and prominent stellar disks started forming only around $7-8$ Gyr ago \citep[see also][]{Park2019}. This has strong implications for the formation and structure of the present-day stellar disk of these galaxies as we will show below. We further note the high degree of symmetry in the oldest stellar components of the galaxies g7.08e11, g7.55e11 and g8.26e11 which persists till $z\sim0.6$ and probably till today.

\subsection{Disk thickness - single vs. double exponential model}

\begin{figure*}
\begin{center}
\includegraphics[width=\textwidth]{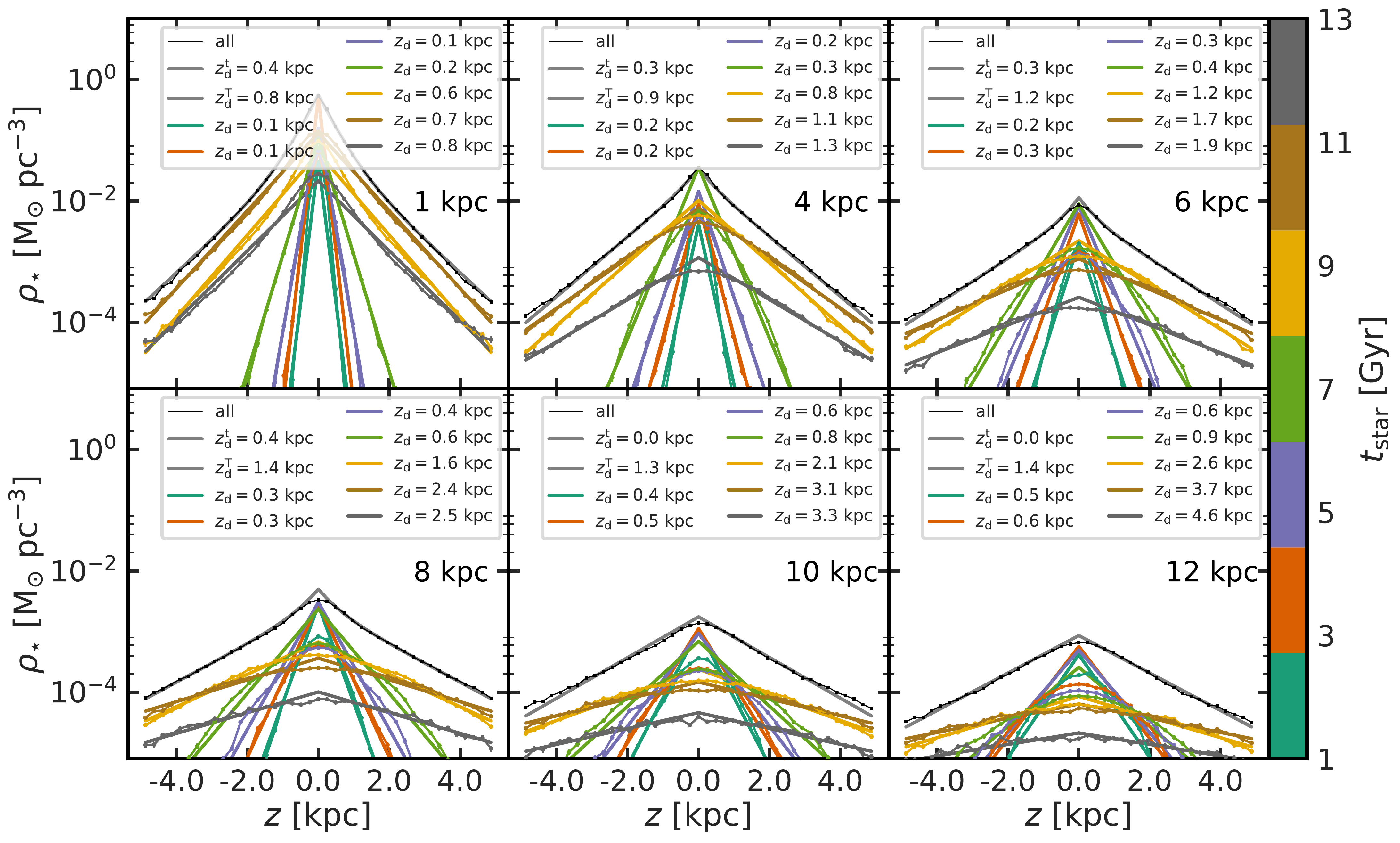}
\end{center}
\vspace{-.35cm}
\caption{Stellar density as a function of height above the stellar disk for different radial bins of width 2 kpc (except for the inner most bin which is $3$ kpc wide) for the galaxy g8.26e11. The bin centroid is indicated in each panel. The black line shows the total stellar density and colored lines split the sample into different mono-age bins as indicated by the colorbar on the right. Exponential fits are shown with dashed lines. Similar figures for all other disk galaxies in the sample are shown in section \ref{app:scaleheight} in the appendix.}
\label{fig:fit}
\end{figure*}

In the following two subsections we evaluate the scale heights as a function of radius and stellar age in order to address the thickness of the disk. We divide the galactic disks into six annuli of width $2$ kpc (except for the inner most bin which we chose to be of $3$ kpc width) and split the star particles into seven mono-age populations in age bins of width $\Delta t=2$ Gyr. For each galaxy we fit the total vertical stellar density with a double exponential reporting the scale heights in table \ref{tab:props}. One example of the resulting fits for galaxy g8.26e11 is shown in figure \ref{fig:fit} where each of the panels shows the different radii bins as indicated by the label. Different colours show the different age bins as indicated in the colorbar on the right hand side of the figure and the black lines in each panel show results for the whole stellar population. Dots show the data with errorbars indicating the uncertainty on the stellar density (mostly smaller than the dot size) and solid lines show the fits to the stellar density. We indicate the measured vertical scale heights for each mono-age population in the legend. A similar figure for all other disk galaxies in the sample is shown in section \ref{app:scaleheight} in the appendix.

Looking at the total vertical stellar density profiles, we find that half of the galaxies are well described by a double exponential while some of the galaxies are better described by a single exponential profile. Galaxies without a unique thick disk have been confirmed by the GHOSTS survey \citep{ghosts}, where several galaxies show single exponential vertical profiles \citep{Streich2016}. Galaxies with clear double exponential profiles show scale heights for the thin and the thick disk components of the order of $h_{\rm z}^{\rm t}\sim200-400$ pc and $h_{\rm z}^{\rm T}\sim1000-1400$ pc, respectively. These scale heights are well in agreement with recent measurements for the MW's stellar disk \citep[see section 5.1.1. in][for a summary of measurements]{Bland-Hawthorn2016} and results from the FIRE group \citep{Ma2017}. Comparing the results for different radii bins (e.g. fig. \ref{fig:fit}), we see that the double exponential vanishes for large radii ($r>9$ kpc).

Focussing on the mono-age populations, we find that each sub-population is well fitted by a single exponential. This is in good agreement with the results from \citet{Martig2014a} who find that both, the radial scale length and vertical scale height of mono age populations in their simulations can be fitted by a single exponential. We especially would like to note that in the simulations where also the total vertical stellar density can be well fitted by a single exponential at all radii, the stellar disk is perturbed by continuous minor mergers like e.g. in the case of g6.96e11. This agrees with previous results studying the impact on mergers on disk heating \citep[e.g.][]{Church2019,Toth1992} and explains why also in the galaxies with double exponential profiles the outskirts are better represented by single exponentials.

Finally, we would like to highlight that this figure as well as the figures in the appendix show that the simulations are very well able to result in exceptionally thin disks of scale heights of $\sim100-400$ pc for the young stars ($<4-5$ Gyr) and $\sim500-1500$ pc for the old stars ($>5$ Gyr). We test the sensitivity of the measured vertical scale height on resolution in the appendix in figure \ref{fig:scale_height_res}. We find that the scale heights in the high resolution simulations are on average $25\%$ smaller compare to the lower resolution with two times worse spatial resolution.

\begin{figure*}
\begin{center}
\includegraphics[width=\textwidth]{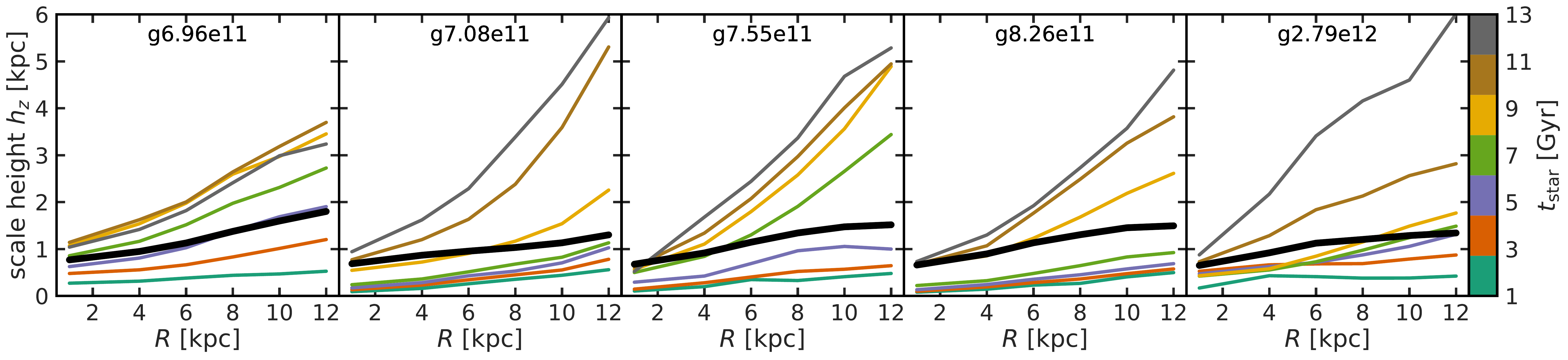}
\end{center}
\vspace{-.35cm}
\caption{Vertical scale height of the stellar disk as a function of radius for all stars in the disk (thick black line) and for different stellar age bins (colored lines). For each individual mono-age bin the stellar vertical scale height increases with radius (except for the youngest age bin in some of the cases). The increase in scale height is much stronger for older mono-age populations while we only see mild increase in the scale height when combining all age populations in good agreement with \citet{Minchev2015}.}
\label{fig:flaring}
\end{figure*}

\subsection{Disk flaring}

Going into detail and looking at each panel in figure \ref{fig:fit} we see that indeed the stellar density for all mono-age populations decreases with radius from left to right and that the scale height is shortest for the youngest stars. Furthermore we see that the scale height for all mono-age populations increases with radius indicating that these galaxies show significant flaring of mono-age populations as proposed by \citet{Minchev2015}. At fixed radius, we find that the vertical scale height increases monotonically with stellar age. We find that a clear double exponential profile results from the relative contribution of young and old stars to the total profile in those galaxies where the scale heights show large differences between two mono-age populations of neighbouring age. This makes us conclude that the NIHAO-UHD simulations \textbf{do not} show a dichotomy between a geometrically thin and thick disk but the scale height of mono-age populations is smoothly decreasing. The double exponential feature as observed in the MW is then simply the result of relative contributions from young and old stars to the total profile. This is in good agreement with previous findings for both the MW \citep[e.g.][although measured from mono-abundance populations]{Bovy2016} or in simulations \citep[e.g.][]{Stinson2013b}.

In figure \ref{fig:flaring} we show the summary of figure \ref{fig:fit} for all disk galaxies in our sample. In each panel we show the scale heights as a function of radius and stellar age with the same color convention as in figure \ref{fig:vel_disp}. For each galaxy we see that each mono-age population of stars flare, i.e. the vertical scale height increases with radius. The strength of the flaring is thereby a strong function of stellar age. Stellar particles younger than $\sim5$ Gyr show only weak flaring while the older populations show strong signatures of flaring. On the other hand, the total scale height of the whole stellar population evolves only weakly with radius as proposed by \citet{Minchev2015}, due to a statistical phenomenon known as Simpson’s paradox \citep[see][]{Minchev2019}. These results are further in agreement with results obtained using the AURIGA galaxies \citep{Grand2017}. The reason for the different scale height evolutions is twofold: First, the older stars show shorter scale length as we show in the next figure such that their density quickly drops with radius and young stars can dominate the scale height at larger radii. Second, the older stars spent more time under the influence of disturbances from close encounters with or even mergers of satellites and subhalos \citep[e.g.][]{Church2019,Toth1992} and secular evolution such as scattering on spiral arms. Thus, they are more prone to vertical heating.

\begin{figure*}
\begin{center}
\includegraphics[width=\columnwidth]{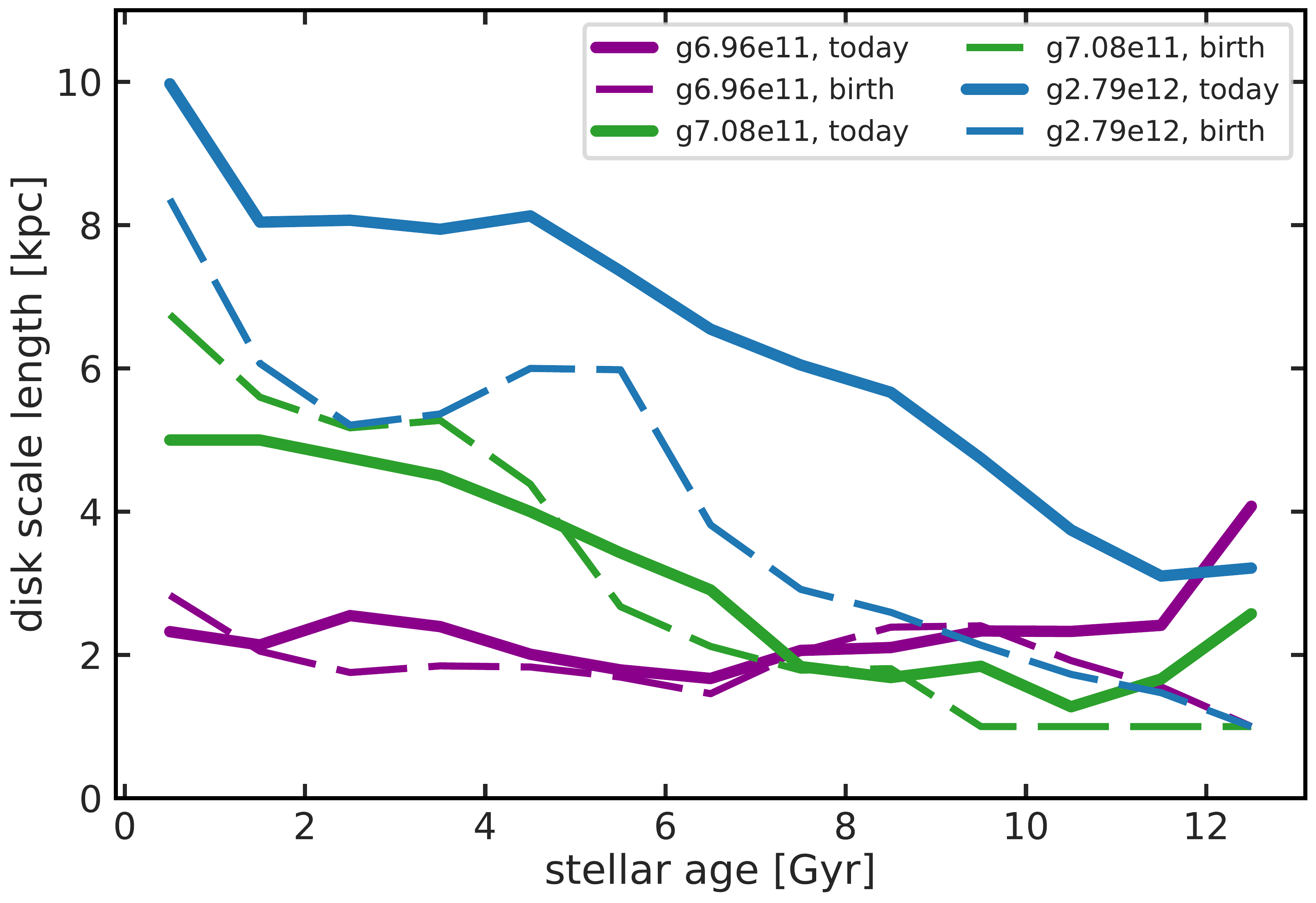}
\includegraphics[width=\columnwidth]{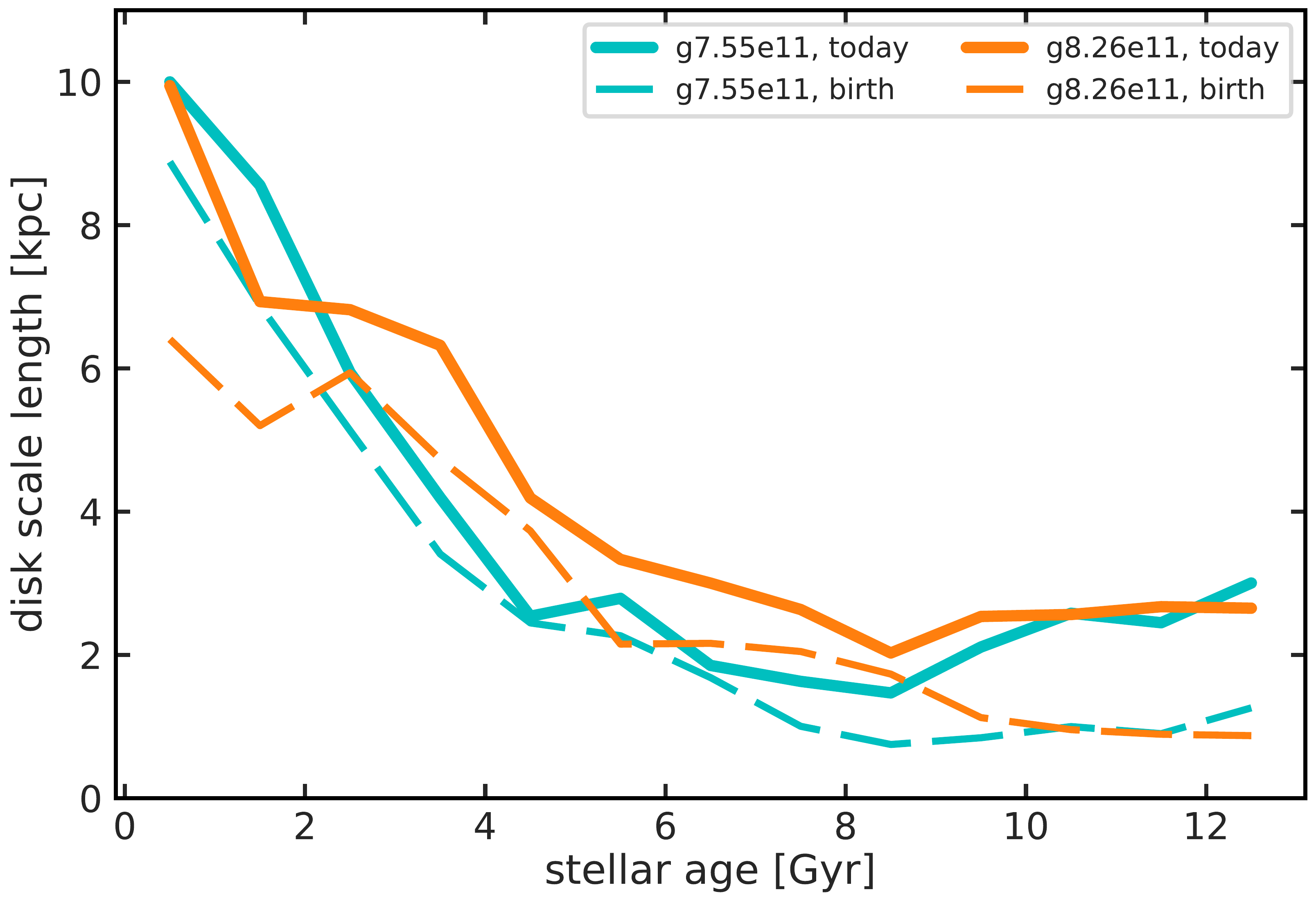}
\end{center}
\vspace{-.35cm}
\caption{Present day radial scale length of disk stars as a function of stellar age (thick lines). For comparison, we also show the scale length of the stars at time of birth with thin dashed lines. For better visibility we split the sample into two panels. We recover the findings from previous figures that older stars show shorter scale length. Further, except for galaxy g6.96e11 and g7.08e11 the scale length for all stellar ages increases strongly with time. This points to a formation scenario of the disk where the scale length of stars is both partly set at the time of birth and partly an effect of subsequent outward migration of disk stars.}
\label{fig:length}
\end{figure*}

\begin{figure*}
\begin{center}
\includegraphics[width=\columnwidth]{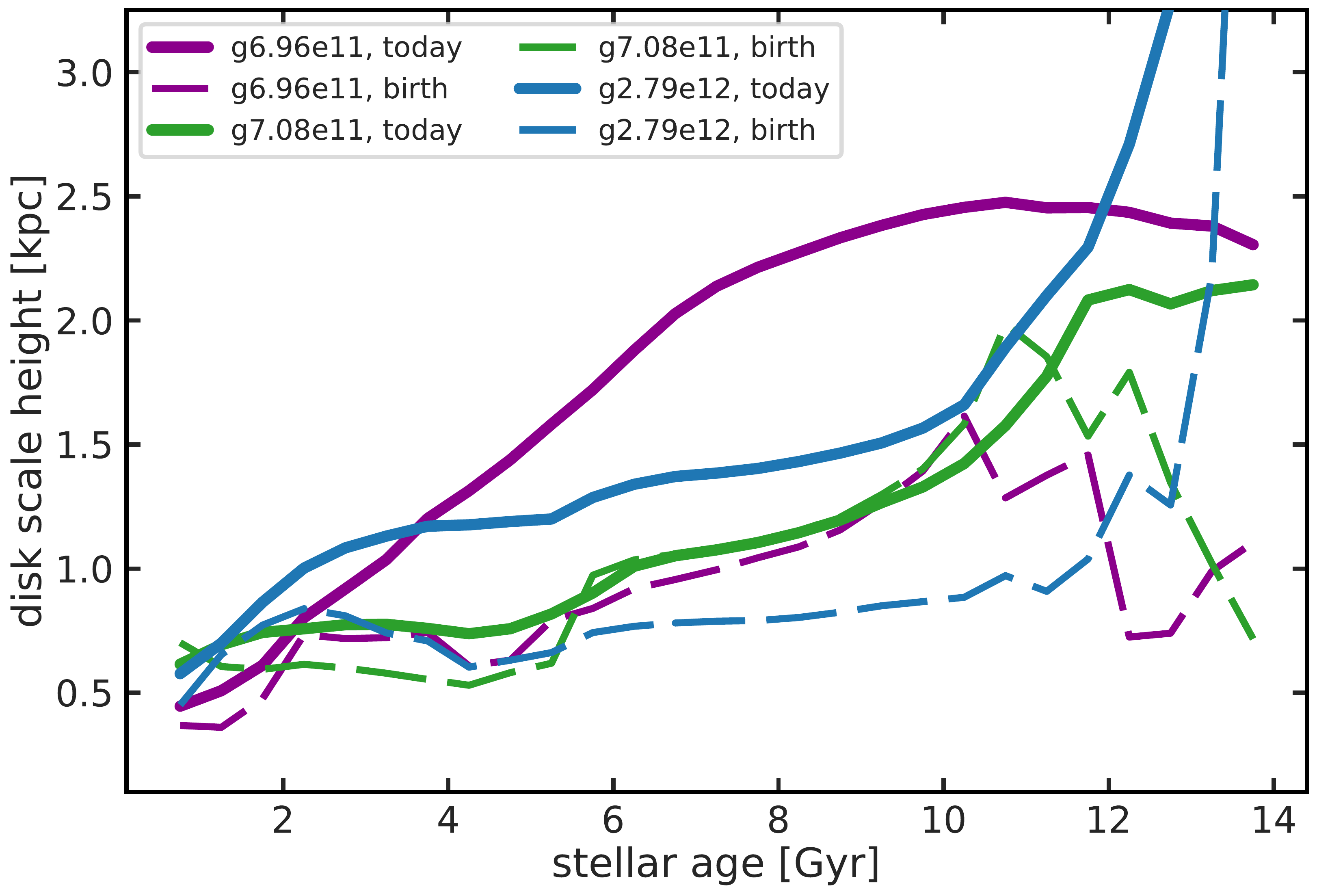}
\includegraphics[width=\columnwidth]{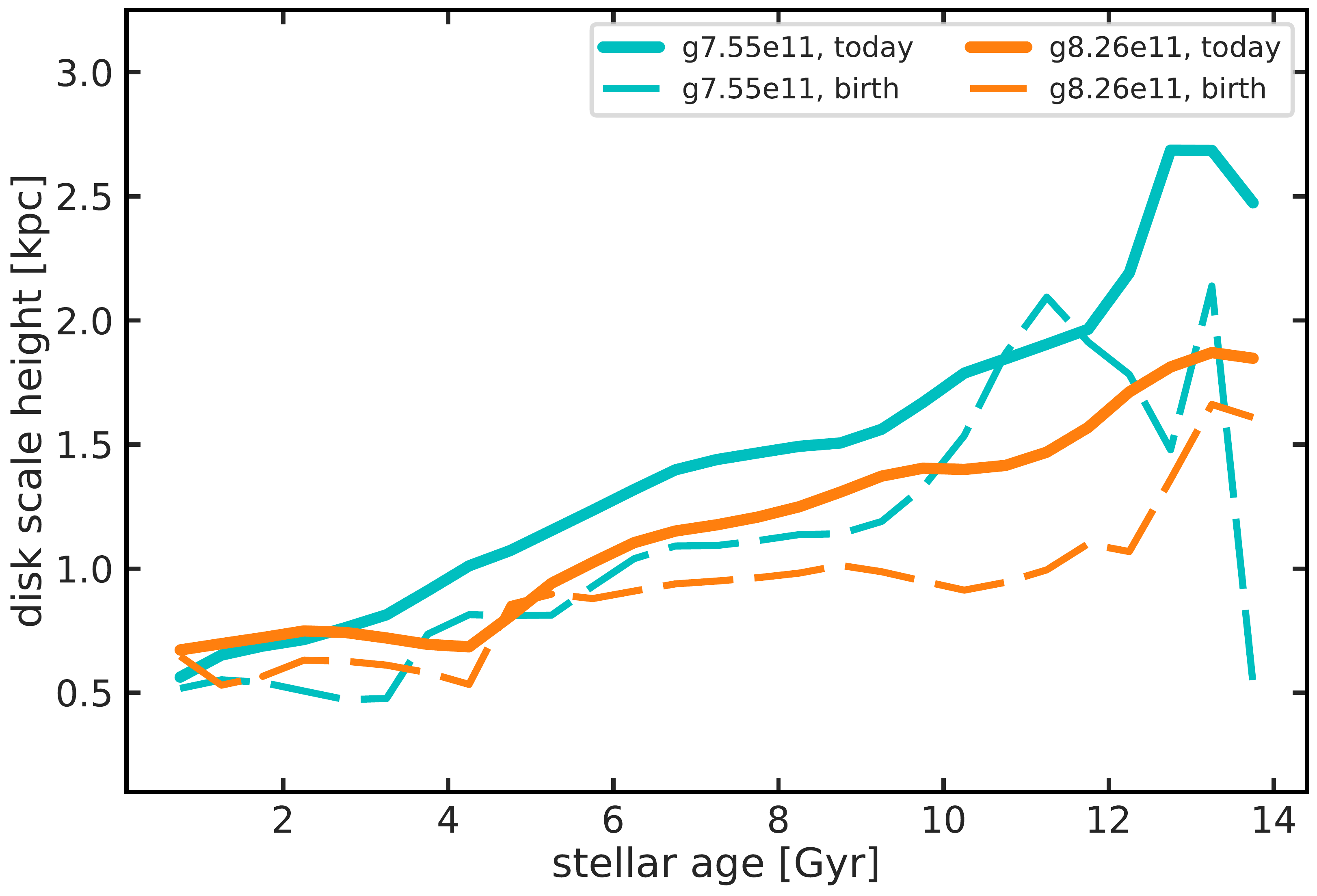}
\end{center}
\vspace{-.35cm}
\caption{Same as figure \ref{fig:length} but for the present day vertical scale height of disk stars as a function of stellar age. For simplicity we neglect here the flaring and select all stellar particles in the radial range $2<R<25$ kpc.   We recover the findings from previous figures that older stars show larger vertical scale heights. Further, except for galaxy g7.08e11 the scale height for all stellar ages increases with time. This points to a formation scenario of the thick disk which is both partly set at the time of birth and partly an effect of subsequent heating of the disk.}
\label{fig:height}
\end{figure*}

\subsection{Time evolution of the scale length and height}

In the next two figures (\ref{fig:length} and \ref{fig:height}) we investigate and quantify the evolution of the scale length and heights of mono-age populations of stars over time in these simulations. Figure \ref{fig:length} shows the scale length of mono-age populations both at present-day (thick lines) and at time of birth (thin dashed lines). For better visibility we split the galaxies into two panels. In figure \ref{fig:height} we show instead the vertical scale height of mono-age populations at present-day and at time of birth. We keep our definition of mono-age population as in the previous figures as stars in age bins of width $\Delta t=2$ Gyr. For simplicity, we selected for this analysis all disk stars within a radius of $25$ kpc but we made sure that the exact choices of radial range does not effect the conclusions made here.

Focussing first on the thick lines in these two figures we find that the scale lengths and scale heights of mono-age populations evolve strongly with stellar age. Younger stars have larger scale length and shorter scale height compared to older age populations. This recovers two results well known as inside-out and upside-down formation of galactic stellar disks \citep{Bird2013,Stinson2013b,Minchev2014} where at early cosmic times stars are born in radially compact but vertically thick disks and then later on when the cosmic accretion rate settles, thin and extended disks are forming.

If we now compare in figure \ref{fig:length} and \ref{fig:height} the present-day (solid colored lines) scale length and heights of each mono-age population to their values at time of birth (thin dashed lines) we find a strong evolution of these properties over time, both scale lengths and scale heights being shorter at time of birth. However, the trends we found at present-day (longer scale length and shorter scale heights for younger stars) are still valid at time of birth. The difference between birth properties and present-day properties is strongest for the galaxies g2.79e12 which has a strong bar and for g6.96e11 which has an extended series of mergers that vertically heated the stellar disk \citep[e.g.][]{Toth1992}. Our findings from fig \ref{fig:length} and fig \ref{fig:height} are in agreement with the results obtained by \citet{Grand2016} using the AURIGA simulations.  

These trends simply reflect the evolution of the gaseous disk out of which the stars form. At early cosmic times the gas is turbulent and thick and stars form only in the central regions of the galaxy with large scale heights or large vertical velocity dispersion. This is also reflected in the results of figure \ref{fig:vel_disp} and figure  \ref{fig:circ} where we found that the vertical velocity dispersion of stars smoothly increases with stellar age and that rotationally supported disks formed only after redshift $\sim2$. With increasing cosmic time, accretion of new material and satellites slows down and the gas settles into a thinner, more extended disk and stars thus form with larger scale length and shorter scale heights. 

However, our analysis shows that this initial structure then gets further modified by secular evolution like vertical heating and radial migration. This has strong consequences for the interpretation of observational findings from MW surveys such as Gaia \citep{Gaia}, Galah \citep{Galah,Buder2018} or 4MOST \citep{4MOST}. Strong secular evolution of the MW's stellar disk implies that detailed models have to be assumed in order to draw conclusions about the formation of the stellar disk from these surveys \citep[for recent models of radial migration in the MW see e.g.][]{Frankel2018,Minchev2018,Feltzing2019}. However, a detailed, observationally motivated study is outside the scope of this paper and will be left for future work.

\section{Summary and discussion}
\label{sec:conclusion}

Reproducing key properties of the MW has been a long standing goal in computational galaxy formation and by now all simulations in the literature successfully reproduce basic properties such as total stellar mass or size \citep[e.g.][]{Wang2015,Grand2017,Hopkins2018}. However, whether these models can simultaneously match also the detailed kinematics or vertical structure of the stellar disk is still not fully settled. Since the first galactic spectroscopic surveys \citep[e.g. RAVE and Gaia-ESO][]{rave,eso} and with the onset of large scale galactic (spectroscopic) surveys like e.g. APOGEE \citep{apogee}, LAMOST \citep{lamost}, Galah \citep{Galah,Buder2018} and finally Gaia \citep{Gaia} which eventually will be complemented and superseded by the 4MOST experiment \citep{4MOST}, our knowledge about the detailed chemo-kinematic structure of the MW's stellar disk has grown tremendously. Still, interpretation of the wealth of data is tricky and often no consensus can be obtained \citep[see discussions in][]{Bovy2016,Duong2018}. 

Part of the problem arises from the fact that slicing the stellar disk in different properties, e.g. either in stellar age, stellar metallicity or [$\alpha$/Fe], results in different patterns and makes a quantitative comparison difficult \citep[e.g.][]{Minchev2017}. This is further complicated by the fact that every Galactic survey is limited to a present-day snapshot of the galactic disk and observed patterns reflect the cumulative effects of birth conditions, secular evolution and cosmological merger history. Thus, complementing observational data of the MW to realistic models of the Galactic growth is inevitably needed in shaping our understanding of the formation of the MW.  

The feedback recipe of the NIHAO simulations results in realistic galaxies across more than two orders of magnitude in halo mass \citep{Wang2015} with baryonic properties matching several key observables of local galaxies \citep[e.g.][]{Obreja2016,Maccio2016,Santos-Santos2018,Roca-Fabrega2019}. Here, we extended the mass resolution of the NIHAO sample roughly a magnitude higher in mass resolution and contrasted our model predictions to observational results obtained for the MW and local massive disk galaxies. We find very good agreement between model predictions and observations in terms of mass, size and kinematics of the stellar and gaseous disks. Convinced by the realism of the model galaxies we further studied the build-up of the stellar disk of these MW-like galaxies across time finding quite some evolution in vertical scale height and radial scale length.

Our results are summarized as follows:
\begin{itemize}
\item \textbf{Stellar mass growth:} The NIHAO-UHD simulation suite results in MW-like galaxies whose stellar masses are in good agreement with abundance matching results across cosmic times (compare fig. \ref{fig:baryon_conv}). At early cosmic times ($z\sim4$) we find that the simulations over-predict the stellar masses compared to the empirical model of \citet{Moster2018}. However, comparing the stellar mass growth as a function of time for the simulations, the empirical model and observations from \citet{vanDokkum2013}, we find that the simulations follow closely the inferred stellar mass growth of MW progenitor galaxies (see fig. \ref{fig:mass_growth}) while stellar masses of star forming galaxies from empirical model are slightly lower at early times. Thus, we conclude that our simulations result in disk galaxies of realistic stellar mass. 

\item \textbf{Accreted stars:} We find that our model galaxies accrete at most a few percent of their present-day stellar mass via mergers, most of them happening early on (dotted lines in fig \ref{fig:mass_growth}) well in agreement with results from \citet{Moster2018}. Thus, our simulations confirm recent findings that significant stellar mergers do not occur for MW like systems. Over $95\%$ of the stars are formed in situ from inflowing gas.

\item \textbf{Stellar disk sizes:}  Stellar disk sizes are characterised by simultaneously fitting a S\'ersic plus exponential profile to the surface density. Resulting disk scale length and effective radii range between $3.9-5.7$ kpc and $0.77-2.96$ kpc, respectively (see fig. \ref{fig:surf_den}). These values are slightly larger than measured for the MW ($R_{\rm d}\sim 3.9$ kpc) but in good agreement with results from the SPARC sample, both as a function of stellar mass and HI properties as figure \ref{fig:sparc} shows. In general, all properties of the simulated galaxies agree well with measurements from the SPARC dataset.

\item \textbf{Rotation curve and stellar kinematics:} Figure \ref{fig:rot} shows that the NIHAO feedback recipe results in flat rotation curves for most of our galaxies. Comparing the rotation velocity as measured at $2.2R_{\rm d}$ or via the HI linewidth reveals that these galaxies further follow the local Tully-Fisher relation (see fig. \ref{fig:tully}).

\item \textbf{Age-velocity dispersion relation:} The vertical velocity dispersion as a function of stellar age for solar neighbourhood stars shows a clear trend of increasing velocity dispersion with age. Comparing the results of the NIHAO-UHD simulations to the observed velocity dispersion of MW and M31 stars in figure \ref{fig:vel_disp} reveals similar trends in our simulations with some of them matching the values of measured velocity dispersion very well. Since this particular diagnostic is a combination of star formation history, birth conditions in the gaseous disk and secular evolution (subsequent heating of the stellar disk plus radial migration of stars) matching the observed values points towards significant realism of the simulations. Adding $25\%$ age uncertainty to the simulations washes out jumps in velocity dispersion due to merger events. This highlights the synergies of a close comparion of simulations and observations to study in detail the build-up of the stellar disk in light of future large scale Galactic surveys.

\item \textbf{Thick or thin disks:} The NIHAO-UHD simulations result in thin and extended stellar disks which with two different kinds of vertical profiles. Half of our sample galaxies have a clear double exponential vertical disk profile while the stellar disk of the other half is well described by a single exponential. The double exponential profile is more pronounced in the central parts ($r\lesssim9$ kpc, e.g. fig \ref{fig:fit}) and we link the occurrence of a single exponential profile to the disk merger history.

\item \textbf{Vertical structure of mono-age populations:}  The vertical density of different mono-age groups of stars are well fitted by single exponential profiles well in agreement with results from e.g. \citet{Mackereth2017}. We find that the vertical scale height increases monotonically with radius at fixed stellar age and similarly with increasing stellar age at fixed radius (e.g. figure \ref{fig:fit}). Thus, we find that the stellar disks, when sliced into mono-age populations, flare (see fig. \ref{fig:flaring}) as is found by \citet{Minchev2015}. Although the total vertical density might be fit by a double exponential, the smooth increase in scale height with stellar age implies that the NIHAO-UHD simulations \textbf{do not} show a distinct geometrically thick disk.

\item \textbf{Formation of the disk:} We investigate what sets the present-day radial and vertical structure of the disk. Figure \ref{fig:length} and fig. \ref{fig:height} show that younger stars tend to have longer radial scale lengths and shorter vertical scale heights both at present-day and at time of birth. Comparing the scale length and height of present-day mono-age stars to their values at time of birth we find that both, the scale length and scale heights of mono-age stars increase strongly with time. If this also holds true for the MW, our results put some tight constraints on how much dynamical memory the stellar disk can retain over cosmological timescales. Detailed comparisons of simulations and observations are needed in order to interpret the wealth of data for Galactic Archeology.

\end{itemize}

\section*{acknowledgments}
We thank the anonymous referee for a careful reading of the manuscript and constructive feedback which helped to improve the quality of the paper.
We further thank Wilma Trick for fruitful discussions and very helpful comments on this work. This research made use of the {\sc{pynbody}} and {\sc{tangos}} package \citet{pynbody,tangos} for simulation analysis and used {\sc{matplotlib}} \citep{matplotlib} to display all figures. Data analysis for this work made intensive use of the {\sc{python}} library {\sc{SciPy}} \citep{scipy}, in particular {\sc{NumPy and IPython}} \citep{numpy,ipython}. This research made use of Astropy,\footnote{http://www.astropy.org} a community-developed core Python package for Astronomy \citep{astropy2013, astropy2018}.
TB acknowledges support from the European Research Council under ERC-CoG grant CRAGSMAN-646955 and from the Sonderforschungsbereich SFB 881 “The Milky Way System” (subproject A2) of the German Research Foundation (DFG). AO is funded by the Deutsche Forschungsgemeinschaft (DFG, German Research Foundation) -- MO 2979/1-1. The authors gratefully acknowledge the Gauss Centre for Supercomputing e.V. (www.gauss-centre.eu) for funding this project by providing computing time on the GCS Supercomputer SuperMUC at Leibniz Supercomputing Centre (www.lrz.de). This research was carried out on the High Performance Computing resources at New York University Abu Dhabi; Part of the simulations have been performed on the DRACO and ISAAC clusters of the Max-Planck-Institut für Astronomie at the Rechenzentrum in Garching. We greatly appreciate the contributions of all these computing allocations.
\bibliography{astro-ph.bib}

\appendix


\section{Resolution study}
\begin{figure*}
\begin{center}
\includegraphics[width=\textwidth]{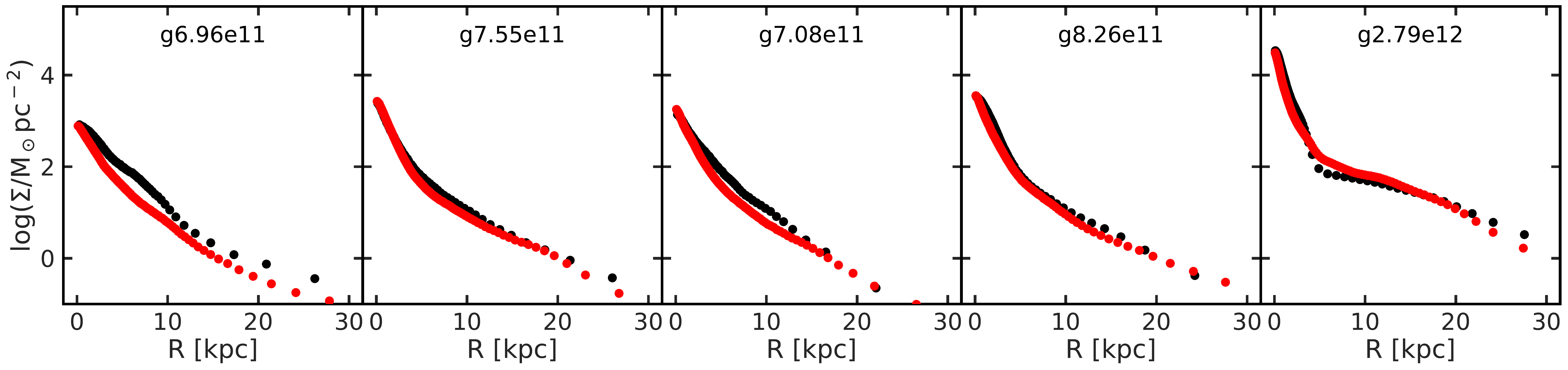}
\end{center}
\vspace{-.35cm}
\caption{Azimuthally averaged stellar surface density profiles for the high (red) and low resolution versions of our model galaxies. Surface densities of different resolution levels agree well.}
\label{fig:surf_res_comp}
\end{figure*}

\begin{figure*}
\begin{center}
\includegraphics[width=\textwidth]{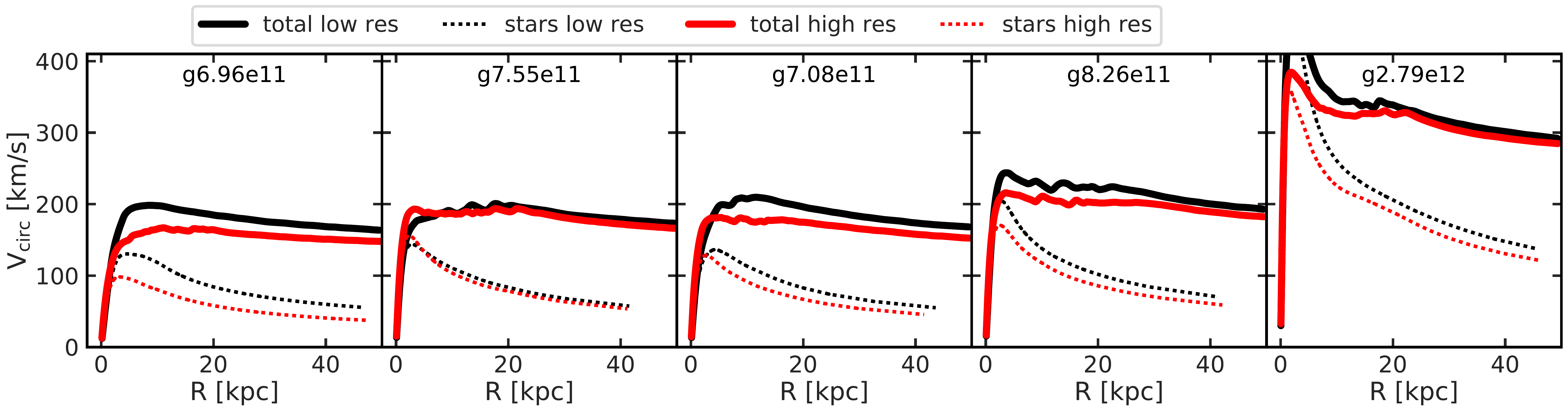}
\end{center}
\vspace{-.35cm}
\caption{Comparison of the rotation curves of high (red) and low resolution (black) model galaxies. Solid lines show the total rotation curve and dotted lines the contribution of the stars. Rotation curves of different resolution levels are reasonably stable.}
\label{fig:rot_curve_comp}
\end{figure*}

\begin{figure}
\begin{center}
\includegraphics[width=\columnwidth]{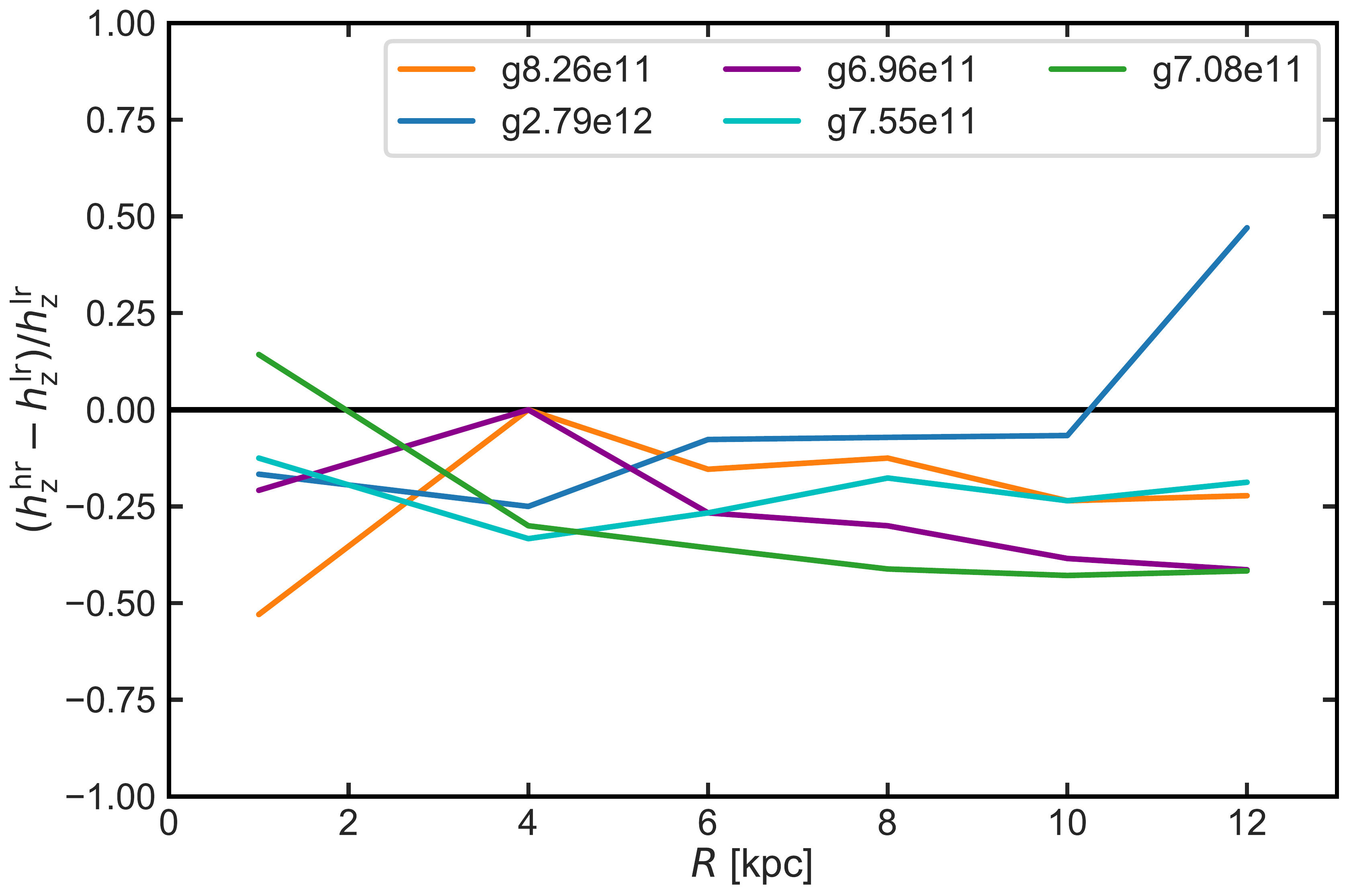}
\end{center}
\vspace{-.35cm}
\caption{Fractional change in scale height of the thick disk between the low- and high-resolution versions of the galaxies. Measured scale heights are on average converged to better than $25\%$.}
\label{fig:scale_height_res}
\end{figure}

In this section we present a resolution study of the main galaxy structural parameters, the stellar surface density (fig. \ref{fig:surf_res_comp}), the rotation curve (fig. \ref{fig:rot_curve_comp}) and the vertical scale height (fig. \ref{fig:scale_height_res}). Unfortunately it is computationally infeasible to run higher resolution simulations thus we compare our results to the lower resolution (two times lower spatial resolution) versions from the NIHAO project. 

Comparing the the surface densities and rotation curves at different resolution levels we find them to be reasonably stable, suggesting that properties which depend on the sub-grid physics such as stellar masses and galaxy sizes are also stable across different resolutions. Comparing the measured vertical scale heights (fig. \ref{fig:scale_height_res}) we find that the high resolution simulations result on average in $25\%$ smaller scale heights. This shows that the scale heights in cosmological simulations are more prone to resolution effects due to the imposed gravitational softenings which prevent the formation of extremely thin disks in low resolution runs. However, the agreement of our high resolution runs with measured vertical scale heights in the MW makes us confident that the employed model successfully produces thin stellar disk of realistic thickness although testing the scale height in even higher resolution is computationally not feasible.

\section{Additional scale height fits}
\begin{figure*}
\begin{center}
\includegraphics[width=\textwidth]{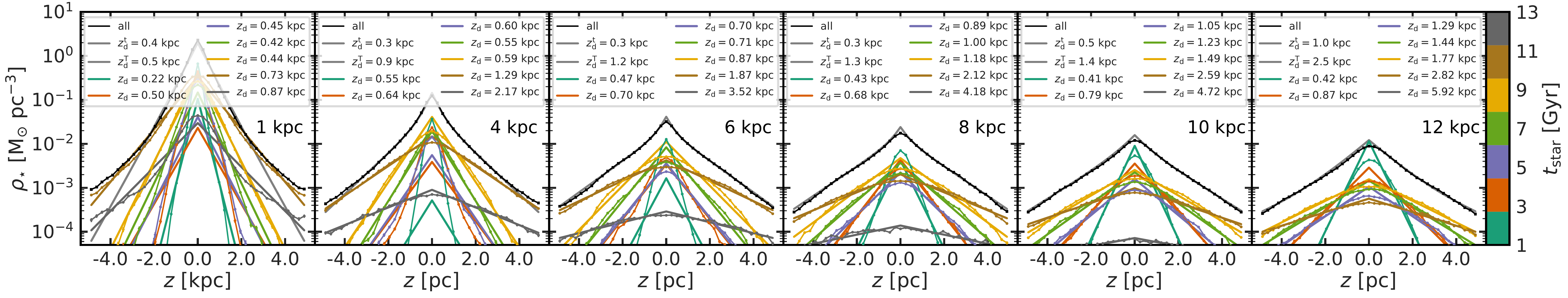}
\includegraphics[width=\textwidth]{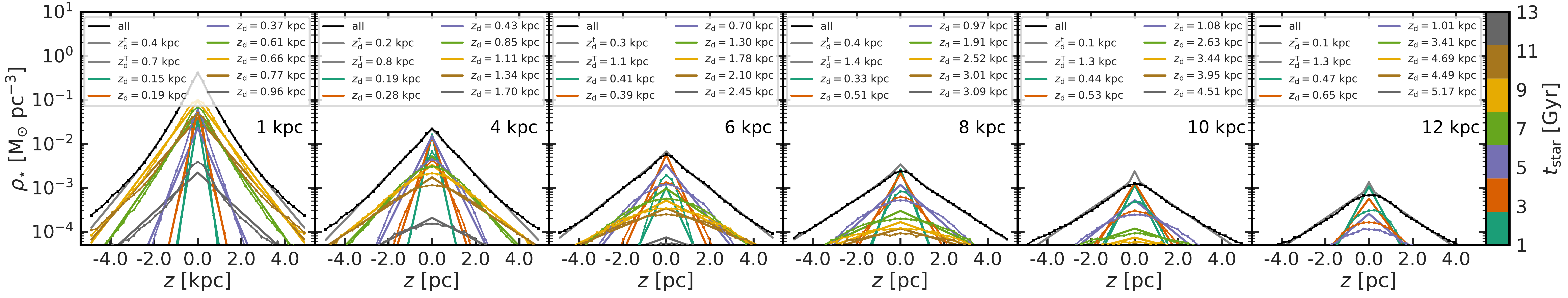}
\includegraphics[width=\textwidth]{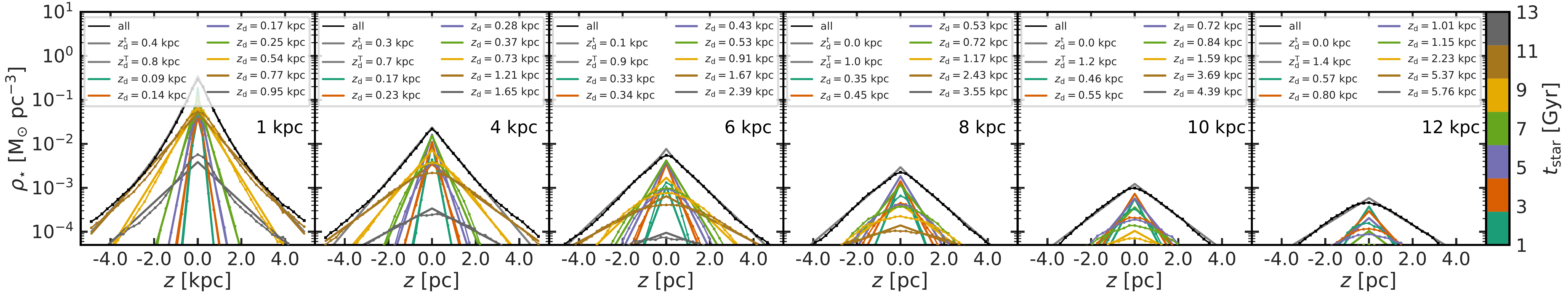}
\includegraphics[width=\textwidth]{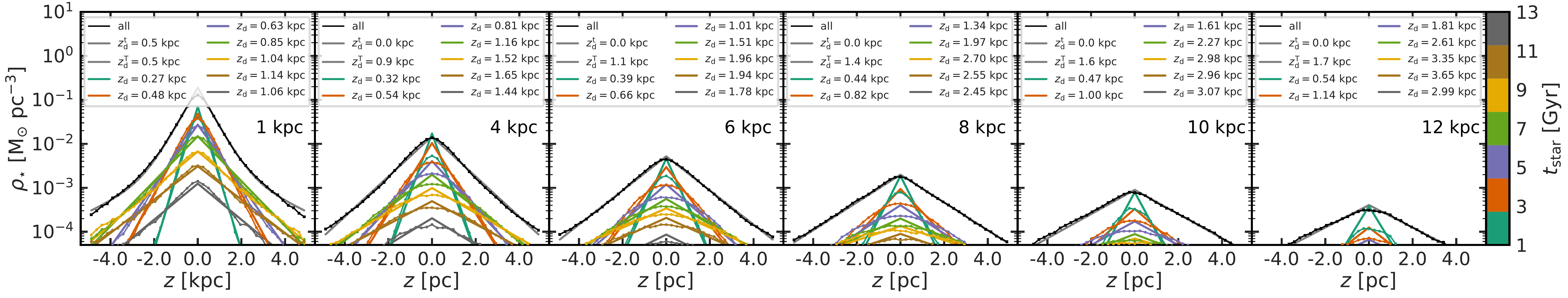}
\end{center}
\vspace{-.35cm}
\caption{Same as figure \ref{fig:height} but for all other disk galaxies in our sample. We show the stellar density as a function of height above the stellar disk for different radial bins of width 2 kpc. The bin centroid is indicated in the upper left corner of each panel. The black line shows the total stellar density and colored lines split the sample into different age bins. From top to bottom we show the galaxies g2.79e12, g7.55e11, g7.08e11, g6.96e11.}
\label{fig:height2}
\end{figure*}

In order to characterise the scale heights of mono-age populations as a function of radius we divide the galactic disks into six annuli of width $2$ kpc and split the star particles into seven mono-age populations of $\pm2$ Gyr. For each of these samples we fit the vertical stellar density with a double exponential although in most cases this reduces to a single exponential fit. In figure \ref{fig:height2} we show from top to bottom the resulting fits for the galaxies g2.79e12, g7.55e11, g7.08e11 and g6.96e11 where from left to right the panels show the different radii bins. In each panel the different colors show the different age bins as indicated in the colorbar on the right. The black line in each panel shows the result for the whole stellar population at this radius bin. Solid lines show the stellar density as measured from the simulation and thin dashed lines show the resulting exponential fits where we indicate the resulting vertical scale heights for each mono-age population in the legend.

\label{app:scaleheight}

\label{lastpage}
\end{document}